\documentclass[aps,superscriptaddress,showpacs,nofootinbib,eqsecnum]{revtex4}

\usepackage{amsmath,lscape,epsfig} \usepackage{amsfonts} \usepackage{amsmath}
\usepackage{amssymb} \usepackage{graphicx} \usepackage{multirow}

\usepackage{slashed}
\usepackage{amsmath,lscape,epsfig}
\usepackage{amsfonts}
\usepackage{amsmath}
\usepackage{amssymb}
\usepackage{graphicx,marvosym}

\usepackage{amsmath,lscape,epsfig} \usepackage{amsfonts}
\usepackage{amsmath} \usepackage{amssymb} \usepackage{graphicx}
\newcommand{\q}{\rm{q}}
\newcommand{\lsim}{\raisebox{-0.13cm}{~\shortstack{$<$ \\[-0.07cm] $\sim$}}~}

\newcommand{\qb}{\overline{\rm{q}}} 
\newcommand{\piv}{\boldsymbol{\pi}}
\newcommand{\pv}{\boldsymbol{p}}
\newcommand{\xv}{\boldsymbol{x}}
\newcommand{\tauv}{\boldsymbol{\tau}}

\newcommand{\lsm}{L$\sigma$M~}
\newcommand{\gammav}{\boldsymbol{\gamma}}
\newcommand{\fls}{\left\langle \Delta^2 \right\rangle}
\newcommand{\flp}{\left\langle \piv^2 \right\rangle}
\def\i{{\rm i}}
\def\uq{{\rm u}}
\def\dq{{\rm d}}
\def\d{{\rm d}}
\def\e{{\rm e}}
\def\Tr{{\rm Tr}}
\begin{document}

\title{Multiple Critical Points in Effective Quark Models} 

\author{Lorenzo Ferroni} \affiliation{Institut f\"{u}r Theoretische Physik, Goethe-Universit\"{a}t, Max-von-Laue St. 1, D-60438 Frankfurt am Main, Germany}\affiliation{Nuclear Science Division,
  Lawrence Berkeley National Laboratory, 94720 Berkeley, CA, USA} 

\author{Volker Koch} \affiliation{Nuclear Science Division,
  Lawrence Berkeley National Laboratory, 94720 Berkeley, CA, USA}

\author{Marcus B. Pinto} \affiliation{Nuclear Science Division,
  Lawrence Berkeley National Laboratory, 94720 Berkeley, CA, USA}
\affiliation{Departamento de Fisica, Universidade Federal de
  Santa Catarina, 88040-900 Florian\'{o}polis, Santa Catarina, Brazil}

\begin{abstract}
 
We consider the two flavor version of the Linear Sigma Model as well 
as of the Nambu Jona-Lasinio model, at finite temperature and quark chemical 
potential, beyond the Mean Field Approximation. Using parameter values for the pion and
quark current masses which weakly break chiral symmetry we show that both models can present more than one
critical end point. In particular, we explicitly show that the appearance of a new critical point associated with a first order line at high temperature and low densities could help to conciliate some lattice results with model predictions. Using different techniques, we perform an extensive thermodynamical analysis to understand the physical nature of the different critical points.  For both models, our results suggest that the new first order line which starts at vanishing chemical potential has a more chiral character than the usual line which displays a character more reminiscent of a liquid-gas phase transition. 

\end{abstract}


\maketitle


\section{Introduction}

Numerical analyses of Quantum Chromodynamics (QCD) on a discrete space-time
lattice (lattice QCD), indicate that the transition from confined to
deconfined matter at finite temperature $T$ and vanishing quark chemical potential $\mu$ 
is a crossover~\cite{Aoki:2006we}. 
On the other hand, model studies~\cite{Asakawa:1989bq,Barducci:1989wi,Barducci:1993bh,Halasz} predict a first-order
transition to occur for $\mu$ of the order of $1/3$ of the baryon mass and 
$T= 0$. In between these two regimes, a second-order critical point is expected 
in the $T-\mu$ plane at some intermediate values of $T$ and $\mu$. 
The existence and the exact location of the critical point is still a matter of dispute~\cite{Fukushima:2008pe} and 
has been under intense theoretical study using effective field theory models of 
QCD~\cite{Asakawa:1989bq,berges98,scavenius,Barducci:1989wi,halasz98,hatta02,Fukushima:2008is,2tri,BoKa} (see also the
recent analysis performed in Ref.~\cite{Zhang:2010qe}). Unfortunately,
a direct application of lattice QCD at finite $\mu$ is, at present, still quite problematic. 
Only relatively recently, new theoretical developments and technical improvements allowed to circumvent 
in various ways the fermion determinant problem and start performing Monte-Carlo calculations 
(see Ref.~\cite{Schmidt:2006us} for a review).  
Although most of the results obtained up to now seem to support the QCD critical point, an interesting 
observation against its existence comes from  Refs.~\cite{deForcrand:2006pv} where, 
from numerical simulations of QCD at imaginary chemical potential, one observes 
that the region of quark masses where the transition is presumably of the first
order (for quark masses smaller than the physical ones), tends to shrink for small positive values of 
the chemical potential, $\mu$. Conversely, according to models supporting the critical point, the first 
order region should expand when $\mu$ increases, so that the physical quark mass point hits the critical 
line at some finite value of $T$ and $\mu$. A possible explanation for this discordance has been given in 
Ref.~\cite{Fukushima:2008is}, where it was pointed out that a  
strong (repulsive) vector coupling may account for the initial shrinkage of the first order region, that 
would then start expanding again at larger values of $\mu$. 
As a result, two critical points might appear for a given range of (small) quark masses, as argued in Ref.\cite {BoKa}. 
However,  it has been remarked ~\cite{Fukushima:2008is} that this 
does not necessarily 
imply the existence of the QCD critical point since a too strong repulsive potential may in fact provoke the disappearance of the first order line (and thus of
the critical point) for physical quark masses. If the vector coupling is too small, instead, the initial 
shrinkage of the first order region is not clearly seen. 
A recent estimate~\cite{fk} of the vector coupling from flavor susceptibilities 
evaluated with lattice QCD seems to support the latter scenario.     

Based on the analysis of
the Linear Sigma Model (L$\sigma$M) with two flavor quarks, it was shown in Refs.~\cite{2tri,BoKa} that  
the inclusion of thermal fluctuations of the mesonic fields leads to the    
appearance of two critical points for a small finite vacuum pion mass, $m_\pi^0< 50$~MeV, 
without the need for a vector interaction.
For physical values of the pion mass the model predicts only one critical point as one would naively 
expect for QCD. Also in this case, the initial shrinkage of the first order region 
at small $\mu$ was proved to be a not uncommon feature. However, as we will discuss,  
the direct transposition of these arguments to QCD should be done with special care.
In the chiral limit, in fact, the phase diagram of the two flavor \lsm is divided in two parts by a 
continuous first order transition, whereas in QCD universality arguments~\cite{Pisarski:1983ms} suggest 
a second order phase transition of the O(4) universality class, at least at $\mu=0$.  
Very recently, it was found (see Ref.~\cite{Skokov:2010sf}) that the correct treatment of the fermion vacuum 
fluctuations (that were neglected in most \lsm applications) can change the order of the transition in the chiral limit 
from first to second order, depending on the coupling constants. In this case the phase diagram 
of the \lsm would resemble the Nambu--Jona-Lasinio model (NJL) one with a second order line 
starting at $\mu=0$ and high-$T$ and terminating at a tricritical point (at intermediate $T$ and $\mu$) where the 
first order line  starts ending at $T=0$.

 
Interestingly enough, these findings of Ref.~\cite{2tri} suggest the possibility of a rich structure for the QCD phase 
diagram in a situation which is similar to the one which arises in metamagnetic systems whose phase 
diagram may display two critical points in the magnetic field versus temperature plane~\cite {wagner}. 
A multicritical point structure induced by charge neutrality and vector interaction has been recently discussed by Zhang and Kunihiro in the context of the 2+1 flavor 
NJL model~\cite{Zhang:2010qe}.
It is worth pointing out that the presence of strong magnetic fields ($B  \simeq 10^{19}\, {\rm G}$), 
may change the order of the phase transition as shown by Fraga and Mizher~\cite{eduana} who considered 
the two flavor \lsm,  at $\mu=0$, obtaining that the usual crossover can turn into a  first order 
phase transition in this regime. 

The aim of the present work is to explore in some details the phase diagram of two 
of the most important effective field theory models of QCD with two quark flavors represented by the \lsm 
and the NJL model when small finite pion (and quark current) masses are considered. The thermodynamics of both models has been compared, with standard 
parametrization 
in Ref.~\cite {scavenius} where the MFA has been used. As we shall see, 
going beyond the MFA allows for the 
appearance of more than one
critical point in both models for certain parameter values.
For the \lsm we will use the same approximation of Refs.~\cite{2tri} 
and  will closely follow the methods therein to map the phase diagram for various values 
of the pion mass, $m_\pi^0$. As it will be shown, by varying $m_\pi^0$, two and also three critical points 
(two of them are actually very close to each other) may appear.
As a further step, we will explore the nature of these critical points by analyzing the susceptibilities 
and the 
correlations of the net-quark number density, the entropy density and the scalar density.

The second part of the paper will be dedicated to the NJL model, in its simplest form, which will be 
treated in the so-called Optimized Perturbation Theory (OPT).  The OPT method (which also
goes by different names, or has many variants,
e.g. ``delta-expansion''~\cite{early}, order-dependent
mapping~\cite{zinn-justin_rev}, etc.)  is well known for allowing evaluations
beyond MFA due to the way it modifies ordinary perturbative expansion, giving
a non trivial (non-perturbative) coupling dependence.   Examples of successful applications include the
most precise analytical values of the critical temperature  for non interacting Bose
gases~\cite{bec} as well as the 
precise location of the tricritical point and the mixed liquid-gas phase 
within the Gross-Neveu
model in 2+1 dimensions~\cite{prdgn3d}. The latter illustrates how this method can be a
powerful tool  beyond standard perturbation theory, since these important
effects were missed by the MFA and could not be precisely determined by Monte
Carlo simulations~\cite{mc}.  
The OPT version adopted here is mainly indicated to non-gauge theories, which
(at finite temperature) require  the method to be  extended, e.g. by adding
and subtracting a hard thermal loop improvement that modifies the propagators
and vertices in a self-consistent way, in the so-called hard-thermal-loop
perturbation theory (HTLpt)~\cite{HTLPT}. Regarding its use within renormalizable theories the OPT 
has just been  substantially improved by its combination with renormalization group properties~\cite {JL}.

This method has been recently applied to the NJL model 
in the evaluation of the thermodynamical potential beyond MFA using standard parametrization ~\cite {prc}. The same type of application  will 
be considered here, with a different set of parameters, in order to verify 
 the possibility of
multiple critical points as found in the L$\sigma$M.  Our investigation, as already anticipated, 
shows that in the  very strong coupling limit this situation (which would be missed by the MFA) arises. 
It is interesting to remark that the OPT brings $1/N_c$ corrections to the MFA  effective potential  
which are proportional to the scalar density, $\rho_s=\langle {\bar \psi} \psi \rangle$, as well as to 
the net-quark number density, $\rho_{\q}=\langle \psi^+ \psi \rangle$, whose  contribution to the pressure 
goes as $-G/(2N_f N_c)(2\rho_{\q}^2-\rho_s^2)$ where $G$ is the usual NJL coupling, $N_c$ is the number 
of colors and $N_f$ the number of flavors.  This means that a type of $1/N_c$ suppressed  vector term   
whose strength is twice its scalar counterpart will contribute to the pressure    so that when the 
interaction is sufficiently strong   the results seem to support the findings of Ref.~\cite{Fukushima:2008is} 
where the SU(3) NJL with an explicit repulsive vector interaction, such as the one suggested in 
Ref.~\cite{koch}, was used.  We recall that, contrary to the L$\sigma$M, the  NJL is a non 
renormalizable theory which is often regularized by a non covariant ultra violet cut 
off\footnote {See Ref.~\cite {scavenius} for an interesting discussion regarding how this 
difference may affect thermodynamical results.}, $\Lambda$. 
Then, from the quantitative point of view our whole NJL application must be taken with care 
since to generate exotic phase diagrams similar to the L$\sigma$M one needs very high values 
for $G$ which in turn generate high effective quark masses at zero temperature and density. 
Although the effective quark mass value generated by those parameters becomes  larger than $\Lambda$, 
the values of relevant observables, such as the quark condensate and the pion decay constant, remain well 
within reasonable values.  Also, as already emphasized, one of the goals of the present NJL 
application is to 
check whether  this model, like the L$\sigma$M, supports the existence of more than one critical point in 
its phase diagram. After obtaining the OPT effective potential (or free energy density) we derive 
the pressure and many thermodynamical quantities of interest including susceptibilities and critical 
exponents. The results obtained with both models  indicate that the first order line observed at 
low chemical potential and high temperature has a more ``chiral'' character while its low temperature 
and high chemical potential counterpart displays characteristics typical of a ``liquid-gas'' phase 
transition. Finally, it will be shown that with both models our results seem to support the 
back-bending of the critical line in the $\mu-m_c$ plane (where $m_c$ is the quark current mass) 
which, as first discussed 
in Refs.~\cite{BoKa,Fukushima:2008is}, could reconcile the actual lattice findings  with most models 
predictions. It is also shown that the MFA completely misses the possibility of more than one 
critical point in the $T-\mu$ plane, for both models.

The paper is organized as follows. In the next Section the L$\sigma$M is reviewed with the 
inclusion of thermal fluctuations. After obtaining the phase diagram in the $T-\mu$ plane for 
small pion masses the characteristics of each different critical point is examined by a 
careful analysis of the densities and susceptibilities. In Sec. III the recent OPT application 
\cite {prc} to the NJL model is quickly reviewed. We then obtain a set of parameter values which 
leads to the emergence of a second critical point, analogous to the one found in the L$\sigma$M.  
We perform a comprehensive thermodynamical analysis to investigate how multiple critical points 
appear in the $T-\mu$ plane as well as in phase coexistence diagrams in the 
$T-\rho_B$ and $P-1/\rho_B$ planes. We also numerically investigate the behavior of quantities 
such as the interaction measure, the equation of state parameter,  bulk viscosity, and susceptibilities 
at, and around, each critical end point. The latter quantities allow us to estimate some relevant critical 
exponents in order to distinguish the physical nature of both critical points which is also done 
by considering the free energy in terms of two ordering densities, $\rho_s$ and $\rho_{\q}$. 
Finally, in  Sec IV we present our main conclusions.

\section{The Linear Sigma Model with quarks}
In standard notation, the density Lagrangian of the \lsm with quarks reads
\begin{equation}
\mathcal{L}=\frac{1}{2}\left(\partial_{\mu} \piv \right)^2+
\frac{1}{2}\left(\partial_{\mu} \sigma \right)^2-U\left(\sigma,\piv\right)
+{\bar \psi}\left[\i \gamma^\mu \partial_{\mu} -g\left(\sigma + \i \gamma_5 \tauv \cdot \piv \right)\right]\psi \; ,
\label{lsm1}
\end{equation}
where $\psi$ is the flavor isodoublet spinor representing the quarks ($\uq$ and $\dq$), and 
\begin{equation}
U\left( \sigma,\piv \right)=\frac{\lambda}{4}\left(\sigma^2+\piv^2-f^2\right)^2-H\sigma \; ,
\label{lsm2}
\end{equation}
is the classical potential energy density. 
In the chiral limit (obtained by setting $H=0$ in the previous equation) the 
chiral symmetry $SU(2)_V\times SU(2)_A$ is spontaneously broken at the classical level, and the pion is the associated
massless Goldston boson.   
For $H \neq 0$, the chiral symmetry is explicitly broken by the term $H\sigma$ in Eq.~(\ref{lsm2})
which gives the pion a finite mass at $T=0$ and $\mu=0$. 
The scalar field $\sigma$ has a finite vacuum expectation value $v$ determined by the classical equation of motion
\begin{equation}
\lambda\left(v^2-f^2\right)v-H=0 \; .
\label{lsm3}
\end{equation}
Accordingly, the $\sigma$ field is conveniently expressed as a sum of the condensate plus fluctuations, 
$\sigma=v+\Delta$. In the \lsm Lagrangian given by Eq.~(\ref {lsm1}) there is no explicit mass term for the quark field, 
the quark mass being 
given only by the condensate, $gv$. 
The parameters of the model are given by the set of equations
\begin{eqnarray}
H&=&f_{\pi}m^{0\;2}_{\pi},\qquad\lambda=\frac{m^{0\;2}_{\sigma}-m^{0\;2}_{\pi}}{2 f^2_{\pi}}, \nonumber \\
f^2&=&\frac{m^{0\;2}_{\sigma}-3m^{0\;2}_{\pi}}{m^{0\;2}_{\sigma}-m^{0\;2}_{\pi}}f^2_{\pi},\qquad g=\frac{m^0_{\q}}{f_{\pi}},
\label{lsm4}
\end{eqnarray}
where $f_{\pi}=92.4$~MeV is the pion decay constant. Consistently with Refs.~\cite{2tri,Mocsy:2004ab} we set
the vacuum $\sigma$ mass to  $m^0_{\sigma}=700$~MeV and the quark mass to $m^0_{\q}=313$~MeV (i.e. one-third of the nucleon mass). 
The last parameter that one needs to fix is the vacuum pion mass, $m^0_{\pi}$, which will be varied from $10$~MeV to $140$~MeV. 

\subsection{Linearized mesonic action for the linear sigma model with quarks}

To calculate the equation of state of the model we closely follow the self-consistent method first proposed
in Ref.~\cite{Mocsy:2004ab} and used also  
in Ref.~\cite{2tri}.
Accordingly, we write the grand canonical partition function as a functional integral in the Euclidean space with 
the imaginary time $\tau=\i t$
\begin{equation}
Z=\rm{Tr}\; \exp\left[-\beta\left(\widehat{H}-\mu \widehat{N} \right)\right]=\int 
\mathcal{D}\psi \mathcal{D}{\bar \psi} \mathcal{D}\sigma \mathcal{D}\piv \exp\left\{\int_{0}^{\beta} \d \tau \int_{V} \d^3 x 
\left(\mathcal{L}+\mu {\bar \psi}\gamma_0 \psi \right)\right\} \; ,
\label{lsm5}
\end{equation}
where $\widehat{H}$ and $\widehat{N}$ are the Hamiltonian and the net quark number operator, respectively.
In Eq.~(\ref {lsm5}), $\beta=1/T$ is the inverse temperature, $\mu$ is the quark chemical
potential\footnote{In this work we use the quark chemical potential $\mu$. The baryochemical potential being
$\mu_{B}=3\mu$.} and $V$ is the system volume. Following the same steps as in Refs.~\cite{2tri,Mocsy:2004ab}, we now integrate away the 
quark degrees of freedom. This amounts to calculate the partition function of the quark sector
\begin{equation}
Z_{\q\qb}=\int 
\mathcal{D}\psi \mathcal{D}{\bar \psi} \exp\left\{\int_{0}^{\beta} \d \tau \int_{V} \d^3 x 
{\bar \psi} \widehat{D} \psi \right\}\,\,,
\label{lsm5.0}
\end{equation}
where 
\begin{equation}
\widehat{D}=-\gamma^{0}\partial_{\tau}+\i \gammav \cdot \nabla-\eta-g\left(\sigma + \i \gamma_5 \tauv \cdot \piv \right)
+\mu \gamma^{0} \; .
\label{lsm5.1pre}
\end{equation}
The Gaussian integral in Eq.~(\ref {lsm5.0}) can be solved with standard techniques~\cite{Kapusta:2006pm} and yields
\begin{equation}
Z_{\q\qb}=\det \widehat{D} \; .
\label{lsm5.1}
\end{equation}
In Ref.~\cite{Aitchison:1985pp} an expression such as Eq.~(\ref {lsm5.1}) has been expanded
in a series of commutators involving the derivatives of the mesonic fields. In our analysis 
we will discard these terms, implicitly assuming that 
the meson mode amplitudes vary slowly in space and time (the same approximation was introduced 
also in Refs.~\cite{2tri,Mocsy:2004ab}) .
The determinant in Eq.~(\ref {lsm5.1}) can then be evaluated as for the free case. 
After performing a Fourier transformation in momentum-frequency space and using the property
\begin{equation}
\ln \det \widehat{D} = \Tr \ln \widehat{D}\;,
\label{lsm5.2}
\end{equation}
we finally obtain
\begin{equation}
\ln \det \widehat{D} = \frac{N_c N_f}{\beta V}\sum_{\pv,n} 
\ln\left\{\beta^2\left[\omega^2_{n} +\left( \varepsilon -\mu \right)^2\right] \right\}
+\ln\left\{\beta^2\left[\omega^2_{n} +\left( \varepsilon +\mu \right)^2\right] \right\}\; ,
\label{lsm5.3}
\end{equation}
where $N_f=2$, $N_c=3$, 
and $\omega_{n}$ are the Matsubara 
frequencies taking the values $\omega_{n}=(2n+1)\pi T$ because of the 
antiperiodicity condition on the fermionic functional integral
$\psi(\xv,0)=-\psi(\xv,\beta)$. In Eq.~(\ref {lsm5.3}) $\varepsilon=\sqrt{p^2+m_{\q}^2}$ is the energy, with 
the quark effective mass
\begin{equation}
m_{\q}^2=g^2\left(\piv^2+\sigma^2\right)\; .
\label{lsm7}
\end{equation}
We note that Eq.~(\ref {lsm5.3}) is formally identical to the standard result except for the 
dependence of $m_{\q}$ on the mesonic fields. Performing the summation over the Matsubara frequencies in 
Eq.~(\ref {lsm5.3}) one then obtains
\begin{eqnarray}
\ln Z_{\q\qb}\left(\sigma,\piv\right)&=&-\int_{0}^{\beta}\d \tau \int_{V} \d^3 x \, \Omega_{\q\qb}\left(\sigma,\piv\right) \nonumber \\
\Omega_{\q\qb}\left(\sigma,\piv\right)&=&-\frac{N_c N_f T}{ \pi^2} \int \d p \; p^2\left\{\beta \varepsilon +\ln\left[1+\e^{-\beta\left(\varepsilon-\mu\right)}\right]+ 
\ln\left[1+\e^{-\beta\left(\varepsilon+\mu\right)}\right] \right\} \; .
\label{lsm8}
\end{eqnarray}
Since we are interested in the low-energy properties of the model, we will
ignore (as was done in~\cite{BoKa,2tri}), for simplicity, the shift in the zero point energy. 
The effect of such a contribution at finite temperature has been analyzed in 
Ref.~\cite{Mocsy:2004ab} for $m_{\pi}=138$~MeV. For physical values of the pion mass, this is not expected to 
change the qualitative behavior of the model.  
As was pointed out in the introduction, however, the 
effect of the fermion vacuum loop, could actually play an important role, changing the order of the 
transition in the chiral limit at $\mu=0$ from first to second order~\cite{Skokov:2010sf}. These contributions
are taken into account in the NJL model where they are responsible of the dynamical breaking of the 
chiral symmetry. 

Using the result in Eq.~(\ref {lsm8})
we can now write an effective Lagrangian that includes only the mesonic degrees of freedom
\begin{equation}
\mathcal{L}=\frac{1}{2}\left(\partial_{\mu} \piv \right)^2+
\frac{1}{2}\left(\partial_{\mu} \sigma \right)^2-U_{eff}\left(\sigma,\piv\right)
\label{lsm9}
\end{equation}
where
\begin{equation}
U_{eff}\left(\sigma,\piv\right)=U\left(\sigma,\piv\right)-\frac{T}{V} \ln Z_{\q\qb}\left(\sigma,\piv\right)\; .
\label{lsm10}
\end{equation}
The Euler-Lagrange equations then read:
\begin{eqnarray}
\partial_{\mu}\partial^{\mu}\sigma+\frac{\partial U_{eff}\left(\sigma,\piv\right)}{\partial \sigma}&=&0 \,\,,\nonumber \\ 
\partial_{\mu}\partial^{\mu}\pi_{i}+\frac{\partial U_{eff}\left(\sigma,\piv\right)}{\partial \pi_{i}}&=&0,\qquad i=1,2,3\;.
\label{lsm11}
\end{eqnarray} 
We now proceed linearizing the mesonic action by taking the average $\langle \ldots \rangle$ of the equations of motion 
over the field fluctuations. First we decompose, as before, the $\sigma$ field as $\sigma=v+\Delta$ where 
$v=\langle \sigma \rangle$  and $\Delta$ is the fluctuation. 
Of course, $\langle \Delta^n \rangle=0$ when $n$ is odd, and the same is true for $\langle \piv^n \rangle$. 
Therefore, since the pion fluctuations always occur as $\piv^2$, the
average $\langle \partial U_{eff}/\partial \pi_{i} \rangle=0$, whereas from the first one of the Eqs.~(\ref {lsm11}) 
we get the condition for the condensate    
\begin{equation}
\left\langle \frac{\partial U_{eff}\left(v+\Delta,\piv\right)}{\partial \Delta} \right\rangle=
\left\langle \frac{\partial U_{eff}\left(v+\Delta,\piv\right)}{\partial v} \right\rangle=0 \; .
\label{eq0}
\end{equation}
The meson masses are then identified with the average of the second derivative of the 
effective potential
\begin{equation}
m^2_{\sigma}=\left\langle \frac{\partial^2 U_{eff}\left(v+\Delta,\piv\right)}{\partial \Delta^2} \right\rangle
\qquad m^2_{\pi}=\left\langle \frac{\partial^2 U_{eff}\left(v+\Delta,\piv\right)}{\partial \pi_i^2} \right\rangle \;,
\label{eq1}
\end{equation}
and the effective potential is linearized as
\begin{equation}
U_{eff}\left(v+\Delta,\piv\right)\sim \langle U_{eff}\left(v+\Delta,\piv\right) \rangle +\frac{1}{2} m^2_{\sigma}
\left(\Delta^2-\fls \right) +\frac{1}{2} m^2_{\pi}
\left(\piv^2-\flp \right)\;.
\label{eq1au}
\end{equation}
The two terms $1/2 m^2_{\sigma} \Delta^2$ and $1/2 m^2_{\pi} \piv^2$ on the right-hand side of the last equation 
are the mass terms to be added to the kinetic
energy in the mesonic Lagrangian to give the mesonic partition function
\begin{eqnarray}
Z_{\sigma ; \piv}&=&\int 
\mathcal{D}\sigma \mathcal{D}\piv \exp\left\{ \frac{1}{2}\int_{0}^{\beta} \d \tau \int_{V} \d^3 x 
\left[  \left(\partial_{\mu} \piv \right)^2+
\left(\partial_{\mu} \sigma \right)^2 -   m^2_{\sigma}
\Delta^2 -   m^2_{\pi} \piv^2  \right]\right\} \nonumber \\
&=& \exp\left[-\frac{V}{T}\left(\Omega_{\sigma}+\Omega_{\pi}\right) \right] \; ,
\label{lsm5m}
\end{eqnarray}
where\footnote{Also for these terms we ignore the shift in the zero point energy, as was done in~\cite{2tri}.}
\begin{eqnarray}
\Omega_{\sigma}&=& \frac{T}{2 \pi^2} \int \d p \; p^2\left[ \frac{1}{2}\beta \varepsilon_{\sigma} +
\ln\left(1-\e^{-\beta\varepsilon_{\sigma}}\right) \right]\,\,,\nonumber \\
\Omega_{\pi}&=& \frac{3 T}{2 \pi^2} \int \d p \; p^2\left[\frac{1}{2} \beta \varepsilon_{\pi} +
\ln\left(1-\e^{-\beta\varepsilon_{\pi}}\right) \right]\; 
\label{lsmq}
\end{eqnarray}
and
\begin{equation}
\varepsilon_{\sigma}=\sqrt{p^2+m^2_{\sigma}}\;,\qquad \varepsilon_{\pi}=\sqrt{p^2+m^2_{\pi}}\;.
\label{mesmass}
\end{equation}
Stemming from Eq.~(\ref {lsm5m}), we have two self-consistency relations between the meson
masses and the corresponding fluctuations
\begin{equation}
\fls=2 \frac{\partial \Omega_{\sigma}}{\partial m^2_{\sigma}}; \qquad 
\flp =2 \frac{\partial \Omega_{\pi}}{\partial m^2_{\pi}} \; .
\label{eq2}
\end{equation}
Finally, we write the thermodynamic potential density 
as\footnote{In the following, where not needed, we will avoid to write explicitly the dependence of the functionals 
on $\sigma$ and $\piv^2$.}
\begin{equation}
\Omega=-\frac{T}{V} \ln Z=\langle U_{eff}\rangle -\frac{1}{2}m^2_{\sigma}\fls-
\frac{1}{2}m^2_{\pi}\flp+\Omega_{\sigma}+\Omega_{\pi} 
\label{gp}
\end{equation}
The average over the field fluctuations is performed by using the techniques of
Refs.~\cite{Mocsy:2004ab,Carter:1996rf,Carter:1998ti}.
Given an arbitrary functional $\mathcal{O}\left(v+\Delta,\piv^2\right)$ we can write down the Taylor expansion 
around $\Delta=\piv^2=0$ and take the average term by term
\begin{equation}
\left\langle \mathcal{O}\left(v+\Delta,\piv^2\right) \right\rangle = \sum_{n,k=0}^{\infty} 
\left. \frac{\partial^{n+k}\mathcal{O}\left(v+\Delta,\piv^2\right)}{\partial \Delta^n\partial \piv^{2k}} 
\right|_{\Delta=\piv^2=0} 
\left\langle \frac{\Delta^n \piv^{2k}}{n!k!}\right\rangle \;.
\label{aver1}
\end{equation}
By using the relation derived in Ref.~\cite{Carter:1996rf}, 
we then relate the terms $\langle \Delta^n \piv^{2k} \rangle$ to 
powers of $\fls$ and $\flp$, i.e. $\langle \Delta^n\rangle=(n-1)!!\fls^{n/2}$ and 
$\langle \piv^{2k} \rangle=(2k+1)!!\langle \piv^2/3 \rangle^k$, which amounts 
of summing up the infinite series of daisy and superdaisy diagrams in the Hartree approximation.
The resulting expression, turns out to 
be equivalent to an integration over a Gaussian distribution~\cite{Mocsy:2004ab}
\begin{equation}
\left\langle \mathcal{O}\left(v+\Delta,\piv^2\right) \right\rangle = 
\int_0^\infty \d z \, P_{\sigma}(z) \int \d y \, y^2 P_{\pi}(y) \mathcal{O}\left(v+z,y^2\right) \; ,
\label{aver2}
\end{equation}
where
\begin{eqnarray}
P_{\sigma}(z)&=&\frac{1}{\sqrt{2 \pi \fls}} \exp\left(-\frac{z^2}{2 \fls}\right) \; , \nonumber \\
P_{\pi}(y)&=&\sqrt{\frac{2}{\pi}}\left(\frac{3}{\flp}\right)^{3/2} \exp\left(-\frac{3y^2}{2 \flp}\right) \; .
\label{aver3}
\end{eqnarray}
In the following, we will need to perform derivatives of the average value of some quantities such as 
the thermodynamic potential density, $\Omega$. After two integration by parts, using Eqs.~(\ref {aver2}) and
Eqs.~(\ref {aver3}) one can obtain the following useful relation~\cite{Mocsy:2004ab}
\begin{equation}
\frac{\partial}{\partial \alpha}\left\langle \mathcal{O}\left(v+\Delta,\piv^2\right)\right\rangle=
\frac{\partial v}{\partial \alpha}
\left\langle \frac{\partial\mathcal{O}\left(v+\Delta,\piv^2\right)}{\partial v}\right\rangle+
\frac{1}{2}\frac{\partial \fls }{\partial \alpha}
\left\langle \frac{\partial^2\mathcal{O}\left(v+\Delta,\piv^2\right)}{\partial \Delta^2}\right\rangle+
\frac{1}{2}\frac{\partial \flp }{\partial \alpha}
\left\langle \frac{\partial^2\mathcal{O}\left(v+\Delta,\piv^2\right)}{\partial \pi_i^2}\right\rangle \; .
\label{average}
\end{equation}
With Eq.~(\ref {average}), and the Eqs.~(\ref {eq0}) and (\ref {eq1}) it is not difficult to see that
\begin{equation}
\frac{\partial \Omega}{\partial v}=\left\langle \frac{\partial U_{eff}}{\partial v}\right\rangle+
\frac{1}{2}\frac{\partial \fls}{\partial v} \left( \left\langle \frac{\partial^2 U_{eff}}{\partial \Delta^2}
\right\rangle -m^2_{\sigma}\right)+\frac{1}{2}\frac{\partial \flp}{\partial v} 
\left( \left\langle \frac{\partial^2 U_{eff}}{\partial \pi_i^2}
\right\rangle -m^2_{\pi}\right)=0 \; .
\label{lsm12}
\end{equation}
In a similar way, using the Eqs.~(\ref {eq2}) one can also easily show that 
\begin{equation}
\frac{\partial \Omega}{\partial m^2_{\sigma}}=\frac{\partial \Omega}{\partial m^2_{\pi}}=0\;.
\label{lsm13}
\end{equation}
Eqs.~(\ref {lsm12}) and (\ref {lsm13}) guarantee the consistency of the approach and the 
standard connection between thermodynamic and statistical mechanics~\cite{Gorenstein:1995vm}.
The equation of state of the system is given by the simultaneous solutions in the variables $v,m_{\sigma},m_{\pi}$ of 
Eqs.~(\ref {eq0}) and (\ref {eq1}), with the field fluctuations given by Eq.~(\ref {lsm13}).
In Ref.~\cite{2tri} this was done numerically and the authors found a rich phase structure with one or 
two critical points, depending on the value of vacuum pion mass $m_{\pi}^0$.
In the next subsection, we will repeat the same calculation as in Ref.~\cite{2tri} and we will show that another 
critical point (very difficult to detect) appears in the phase diagram. 

\subsection{The phase diagram}

We now proceed and map the phase diagram of the model for various values of the vacuum pion mass ranging from 
$m_\pi=10$~MeV to $m_\pi=140$~MeV. 
As a first step, we need to find a way to localize the first order line(s) and the critical 
end point(s). This is usually done by looking at the change in shape of the thermodynamic potential 
across a transition line. 
\begin{figure}[h!]
\begin{center}
\includegraphics[width=0.42\textwidth]{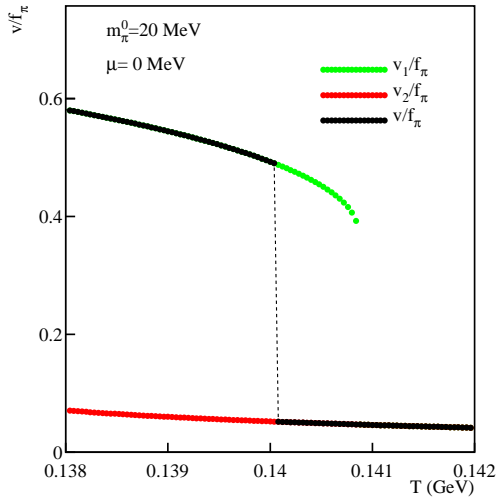}
\caption{\small{(Color online) The condensate $v$ (normalized to $f_{\pi}$) as a function of the temperature 
in the vicinity of the first order 
transition for $m^0_{\pi}=20$~MeV and $\mu=0$. The stable solution $v$ (shown with black markers and the dashed line) is the 
one that corresponds to the highest pressure. The other two (metastable) solutions $v_1$ and $v_2$ are also shown in the
picture.}}
\label{dsol}
\end{center}
\end{figure}
This method (that will be adopted for the NJL model in sec.~\ref {secnjl}), 
allows to see quite clearly the onset (and usually the order) of a phase transition, but
unfortunately cannot be used for the present analysis. Our approximation is, in fact, based on an expansion 
of the thermodynamic potential around a minimum and our equations are defined only there so that a 
different method must be employed. 
A typical signature of the onset of a first order transition is the presence of two
minima of the thermodynamic potential, corresponding to the high and low temperature phases. 
In turn, we should see two distinct solutions for $v,m_{\sigma},m_{\pi}$ at the same $T$ and $\mu$. 
The transition happens when these two solutions have the same potential (pressure), i.e. the point when the 
system switches from one minimum to the other.
For each value of $\mu$, we solve the system of equations starting from low temperatures 
(the broken phase\footnote{Since the symmetry is explicitly broken 
by the vacuum pion mass, expressions such as ``symmetric phase'' or ``broken phase'' must
be understood just as a nomenclature convention.}) and following the line of minima 
increasing the temperature by a small amount $\Delta T$ at each step, using the latest solution as initial point for the
solving routine. 
Once we are sure to be in the high $T$ symmetric phase, we solve the system of equations 
backwards (going from high to small $T$) until we are sure to be below the 
transition temperature. If the transition is continuous, we will find a unique solution for each value of $T$.
Conversely, if the transition is of the first order, we will have a region where two solutions exist at the same $T$.

In Fig.~\ref {dsol} this is shown for the condensate $v$ around the transition line for $m^0_{\pi}=20$~MeV.
For this value of the pion mass, the system undergoes a first order transition at $T\sim 140$~MeV and
$\mu=0$. As one can see, the solution corresponding to the symmetry-broken phase ($v_1$) exists at low $T$ and shortly
above the transition temperature, whereas the symmetric solution ($v_2$) exists at high $T$ and below the transition.
The actual solution (the one corresponding to the bigger pressure) is shown by black dots and the dashed line.

\begin{figure}[h!]
\begin{center}
\includegraphics[width=0.6\textwidth]{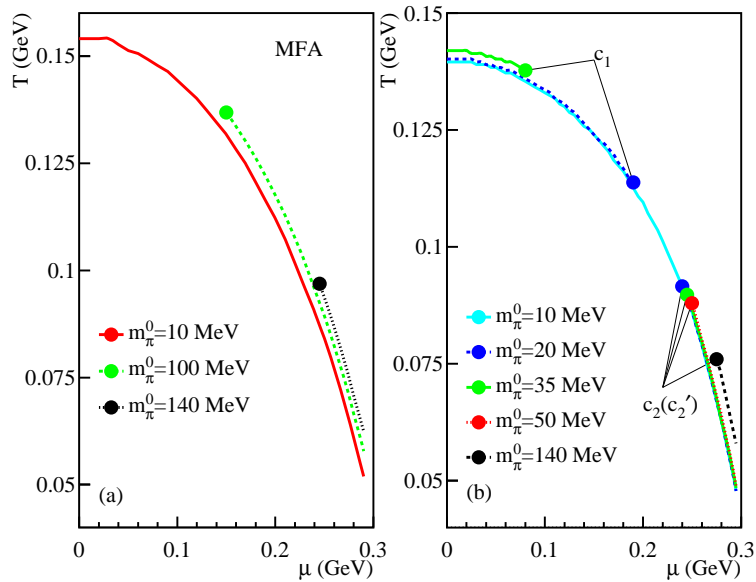}
\caption{\small{(Color online) Left panel: the phase diagram of the \lsm in the MFA 
for $m^0_{\pi}=10,100,140$~MeV. Right panel: 
the phase diagram of the \lsm with the inclusion of mesonic fluctuations for 
$m^0_{\pi}=10,20,35,50,140$~MeV}}
\label{f3}
\end{center}
\end{figure}

With this method, we now map the phase diagram of the model in the $T-\mu$ plane.
Our results are consistent with those in Ref.~\cite{2tri}, where the authors performed the same calculation and found
that for sufficiently small values of the vacuum pion mass ($m^0_{\pi}\lesssim 50$~MeV), the system has two critical 
points.   
In Fig.~\ref {f3} we plot the phase digram of the model in the MFA (Fig.~\ref {f3}a) and with the inclusion of mesonic
fluctuations (Fig.~\ref {f3}b).
In Fig.~\ref {f3}b we have analyzed various values of $m^0_{\pi}=10,20,35,50,140$~MeV. 
For $m^0_{\pi}=10$~MeV the phase diagram is divided in two parts by a continuous first order line. 
For $m^0_{\pi}=20$~MeV and $35$~MeV the first order line is interrupted by a continuous region, and
for some $35 < m^0_{\pi} \leq 50$~MeV the branch at low $\mu$ disappears and we have the usual continuous 
transition at low $\mu$ and first order at high $\mu$.

\begin{figure}[h!]
\begin{center}
\includegraphics[width=0.47\textwidth]{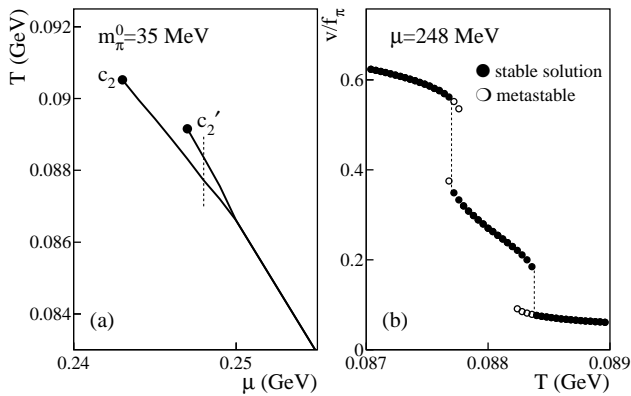}
\caption{\small{Panel a: a detail of the phase diagram for $m^0_{\pi}=35$~MeV showing
the rightmost first order line bifurcating and ending in two critical points. Panel b:
the condensate $v$ (normalized to $f_{\pi}$) as a function of the temperature for $\mu=248$~MeV. The plotted region 
corresponds to the dashed vertical line crossing the 
two first order lines in Panel a. The stable solution is represented by black dots and the metastable
solutions by open circles.}}
\label{f4}
\end{center}
\end{figure}

The leftmost branch of the first order line (when it exists) ends in a critical point.
For the rightmost one, instead, the situation seems to be different.
A closer look revealed the presence of a double critical point.
This is shown in Fig.~\ref {f4}a for $m^0_{\pi}=35$~MeV. As one can see, the first order line
bifurcates and ends in {\em two} critical points that have been labeled $c_2$ and $c_2'$.
Their existence is revealed by the presence of two 
first order transitions that have been detected, as before, looking at the double solutions of our system of equations.
In Fig.~\ref {f4}b we show the parameter $v$ along the dashed line crossing the two first order lines in 
Fig.~\ref {f4}a. As one can see, there are two distinct solutions (that we have checked to correspond to minima 
of the thermodynamic potential) for two slightly different ($\sim 1$~MeV) values of the temperature, 
indicating a double first order. 
\begin{figure}[h!]
\begin{center}
\includegraphics[width=0.4\textwidth]{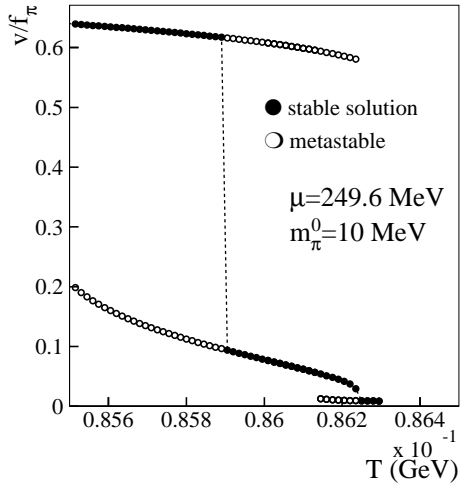}
\caption{\small{The condensate $v$ (normalized to $f_{\pi}$) as a function of the temperature for $\mu=249.6$~MeV 
and $m^0_{\pi}=10$~MeV. 
Two first order transitions are visible at $T\sim85.9$~MeV and $T\sim86.2$~MeV.
The stable solution is represented by black dots and the metastable
solutions by open circles.}}
\label{ufine}
\end{center}
\end{figure}
A similar behavior can be observed for all the explored values of the pion mass, i.e. $20,35,50,140$~MeV. 
Also for $m_{\pi}=10$~MeV, where the transition is of the first order everywhere, a very short appendix 
of the the first order line appears around $\mu\sim 250$~MeV and $T\sim 86$~MeV producing
another very short branch ending in a critical point 
at $\mu\sim 249$~MeV and $T\sim 87$~MeV. The situation is qualitatively similar to the one in Fig.~\ref {f4}a except 
for the fact that 
the leftmost critical point $c_2$ does not exist as the first order line continues until $\mu=0$.
In Fig.~\ref {ufine} one can see the multiple solutions for $v$, 
indicating the two first order lines at $T\sim85.9$~MeV and $T\sim86.2$~MeV, respectively. 
For the rightmost transition, the 
discontinuity in $v$ is very small, but its presence is proved by the existence of the
metastable solution. 
At present the reason for the this
unusual bifurcation of the  high-$\mu$ first order line is not
understood and requires further studies. It may simply be an artifact
of the approximation (mean-field plus fluctuations) used for the treatment of the linear sigma
model. However, as we will see in the next section, despite their
vicinity in the phase digram, the two first order lines appear to have slightly different qualitative features.

\subsection{Densities and susceptibilities}
We now go one step further and analyze the
susceptibilities in the \lsm with the aim of characterizing the critical point(s) by
studying the fluctuations of the net-quark number density $\rho_{\q}$, scalar density $\rho_{s}$, and entropy density 
$s$ in the vicinity of the critical regions. 
By definition, the quark density is  
\begin{equation}
\rho_{\q}=\frac{T}{V} \frac{\partial \ln Z}{\partial \mu}=-\frac{\partial \Omega}{\partial \mu}\;.
\label{qdens1}
\end{equation}
By using the Eq.~(\ref {average}) and Eqs.~(\ref {lsm12}) and (\ref {lsm13}) one then has
\begin{equation}
\rho_{\q}=-\left\langle \frac{\partial U_{eff}}{\partial \mu}\right\rangle=
\frac{T}{V} \left\langle \frac{\partial \ln Z_{\q\qb}}{\partial \mu}\right\rangle \;.
\label{qdens2}
\end{equation}
To calculate the scalar density, it is convenient to introduce a mass term $-\eta
{\bar \psi}\psi$ for the quarks in the Lagrangian in Eq.~(\ref {lsm1}), where $\eta$ is a fictitious bare quark mass to be set to 
zero afterwards. This amounts to a thermal 
quark mass
\begin{equation}
m_{\q}^2=g^2\piv^2+\left(g\sigma + \eta\right)^2 \; .
\label{tqm2}
\end{equation}
From the partition function written as in Eq.~(\ref {lsm5}) one then sees that
\begin{equation}
\rho_{s}= \langle {\bar \psi}\psi\rangle =
\left. -\frac{T}{V} \frac{\partial \ln Z}{\partial \eta} \right|_{\eta=0} = 
\left. \frac{\partial \Omega}{\partial \eta} \right|_{\eta=0}\;, 
\label{sdens1}
\end{equation}
that is, as before,
\begin{equation}
\rho_{s}
=\left\langle  \frac{\partial U_{eff}}{\partial \eta}  \right\rangle_{\eta=0}
= -\frac{T}{V} \left\langle \frac{\partial \ln Z_{\q\qb}}{\partial \eta}\right\rangle_{\eta=0} \; .
\label{sdens2}
\end{equation}
Unlike the quark density in Eq.~(\ref {qdens2}), the scalar density can be directly evaluated, once $v$ and the meson
masses are known, without the need to solve any further integral. Indeed, from Eq.~(\ref {eq0}) one 
gets
\begin{equation}
\frac{T}{V} \left\langle \frac{\partial \ln Z_{\q\qb}}{\partial v} \right\rangle=\lambda \left(v^2 + 3\fls+\flp -f^2\right)v-H \; .
\label{sdens3}
\end{equation}
Noting that  
\begin{equation}
\frac{\partial \ln Z_{\q\qb}}{\partial \eta} = \frac{\partial \ln Z_{\q\qb}}{\partial v} 
\frac{\partial m^2}{\partial \eta} \left(\frac{\partial m^2}{\partial v}\right)^{-1}=
\frac{1}{g}\frac{\partial \ln Z_{\q\qb}}{\partial v}
\label{sdens4}
\end{equation}
and using Eqs.~(\ref {sdens2}) and (\ref {sdens3}) we finally obtain
\begin{equation}
\rho_{s}=\frac{1}{g }\left[H-\lambda \left(v^2 + 3\fls+\flp -f^2\right)v\right] \;  .
\label{sdens5}
\end{equation}
The net-quark number density and the scalar density are the thermodynamic variables conjugate to $\mu$ and the fictitious
quark mass $\eta$, respectively. The last density that we need to evaluate is the the entropy density (conjugate to the
temperature $T$), i.e.
\begin{equation}
s=-\frac{\partial \Omega}{\partial T}=\left\langle\frac{1}{V}\ln Z_{\q\qb} + \frac{T}{V} \frac{\partial \ln Z_{\q\qb}}{\partial T} \right\rangle 
-\frac{\partial \Omega_{\sigma}}{\partial T}-\frac{\partial \Omega_{\pi}}{\partial T} \; .
\label{edens1}
\end{equation} 
In what follows, we will always assume the same number of $\uq$ and $\dq$ quarks in the system. 
The isospin density is therefore always vanishing 
for any value of $T$, $\mu$ and $\eta$, and, of course, the same is true for its derivatives with respect to 
these quantities. 
The Isospin density, therefore, do not mix with the other degrees of freedom\footnote{Note that even though 
the correlations with the other densities vanish, the Isospin susceptibility 
(i.e. the derivative of the Isospin with respect to the Isospin chemical potential) do not.} 
($\rho_{\q}$, $\rho_s$ and $s$) and will be not considered in our analysis at this time.

\begin{figure}[h!]
\begin{center}
\includegraphics[width=0.47\textwidth]{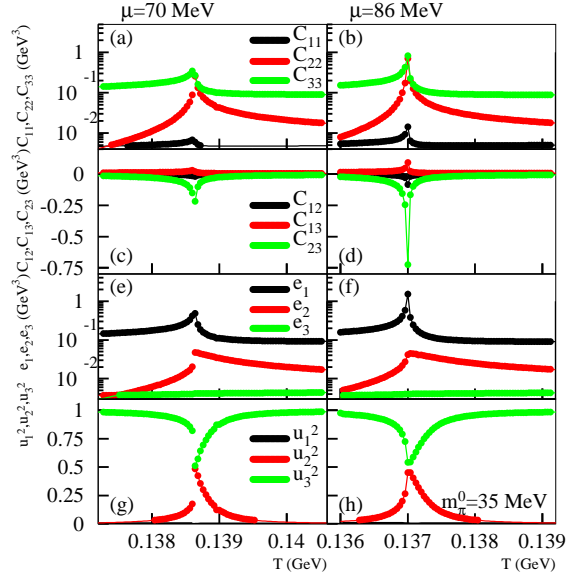}
\caption{\small{(Color online) Panels a and b: diagonal matrix elements of the covariance matrix as a function of the 
temperature in the vicinity of
the critical point $c_1$ for $m^0_{\pi}=35$~MeV and $\mu=70$~MeV and $\mu=86$~MeV, respectively.
Panels c and d: off-diagonal matrix elements of the covariance matrix as a function of the temperature in the vicinity of
the critical point $c_1$ for $m^0_{\pi}=35$~MeV and $\mu=70$~MeV and $\mu=86$~MeV, respectively.
Panels e and f: eigenvalues of the covariance matrix as a function of the temperature in the vicinity of
the critical point $c_1$ for $m^0_{\pi}=35$~MeV and $\mu=70$~MeV and $\mu=86$~MeV, respectively.
Panels g and h: square components of the (normalized) eigenvector corresponding to the larger eigenvalue ($e_1$) of the covariance
matrix as a function of the temperature in the vicinity of
the critical point $c_1$ for $m^0_{\pi}=35$~MeV and $\mu=70$~MeV and $\mu=86$~MeV, respectively.}}
\label{s1}
\end{center}
\end{figure}
\begin{figure}[h!]
\begin{center}
\includegraphics[width=0.47\textwidth]{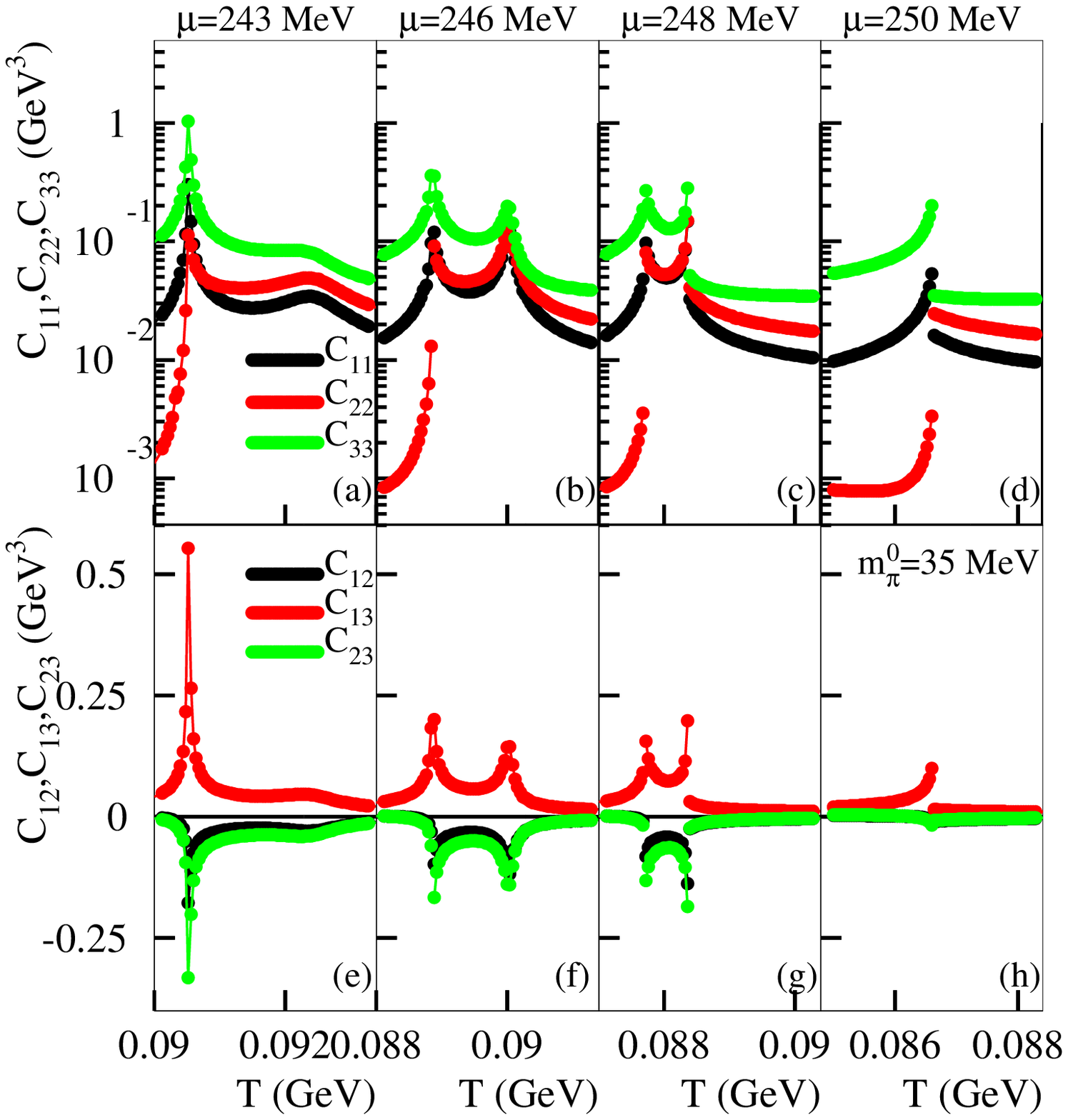}
\caption{\small{(Color online) Panels a, b, c and d: diagonal matrix elements of the covariance matrix 
in the vicinity of
the critical points $c_2$ and $c_2'$ as a function of the temperature for $m^0_{\pi}=35$~MeV 
and $\mu=243$~MeV,$\mu=246$~MeV, $\mu=248$~MeV and $\mu=250$~MeV, respectively.
Panels e, f, g and h: off-diagonal matrix elements of the covariance matrix 
in the vicinity of
the critical points $c_2$ and $c_2'$ as a function of the temperature for $m^0_{\pi}=35$~MeV 
and $\mu=243$~MeV,$\mu=246$~MeV, $\mu=248$~MeV and $\mu=250$~MeV, respectively.}}
\label{s2}
\end{center}
\end{figure}
We can now go ahead and calculate the $3 \times 3$ covariance matrix as
\begin{equation}
 \mathbf{C} = T \left(
  \begin{array}{ccc}
   \frac{\partial \rho_{\q}}{\partial \mu} & -\frac{\partial \rho_{\q}}{\partial \eta} &  \frac{\partial \rho_{\q}}{\partial T} \\
 \frac{\partial \rho_{s}}{\partial \mu} & -\frac{\partial \rho_{s}}{\partial \eta} &  \frac{\partial \rho_{s}}{\partial T} \\  
 \frac{\partial s}{\partial \mu} & -\frac{\partial s}{\partial \eta} &  \frac{\partial s}{\partial T}
  \end{array} \right) \equiv \mathbf{C}_{ij} \qquad i,j=1,2,3 \; .
\label{cmatr}  
\end{equation}
This matrix is symmetric and can be diagonalized. We call $e_1, e_2, e_3$ the three eigenvalues of $\mathbf{C}$
and $\bf{u},\bf{v},\bf{z}$ the corresponding eigenvectors.
Each one of the three eigenvectors can be related to a different orthogonal 
combination of the three original densities $\rho_{\q},\rho_{s}$ and $s$
\begin{eqnarray}
\rho_{\bf u}&=&u_1\rho_{\q}+u_2\rho_{s}+u_3 s \nonumber \\
\rho_{\bf v}&=&v_1\rho_{\q}+v_2\rho_{s}+v_3 s\nonumber \\
\rho_{\bf z}&=&z_1\rho_{\q}+z_2\rho_{s}+z_3 s \; .
\label{divdens}  
\end{eqnarray}
These three new densities are now independent. At the critical point, only one of them 
will have divergent fluctuations, whereas the other two will remain finite.
We chose, as a convention, to order the eigenvalues from the biggest $e_1$ to the smallest $e_3$.
At the critical point, we then expect the first eigenvalue, 
$e_1$, to diverge. According to Eq.~(\ref {divdens}), the three components of the corresponding 
eigenvector ${\bf u}=\left(u_1,u_2,u_3\right)$ 
will then give us the expression of the critical density in terms of the original densities $\rho_{\q},\rho_{s}$ and $s$. 
\begin{figure}[h!]
\begin{center}
\includegraphics[width=0.47\textwidth]{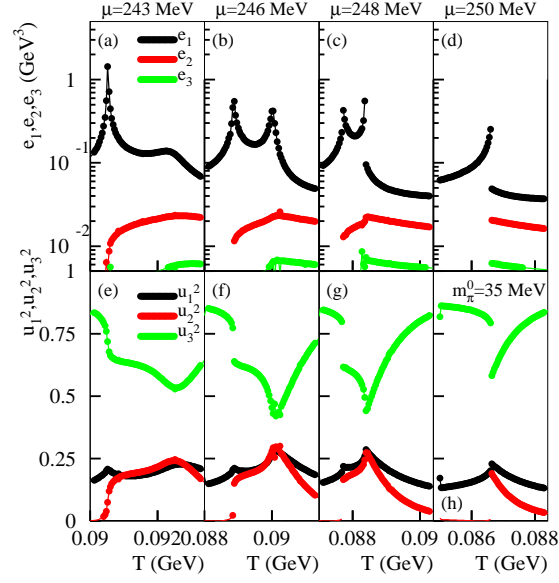}
\caption{\small{(Color online) Panels a,b,c and d: eigenvalues of the covariance matrix as a function of the temperature in 
the vicinity of the critical points $c_2$ and $c_2'$ for $m^0_{\pi}=35$~MeV and 
$\mu=243$~MeV,$\mu=246$~MeV, $\mu=248$~MeV and $\mu=250$~MeV, respectively.
Panels e,f,g and h: square components of the (normalized) eigenvector corresponding to the larger eigenvalue ($e_1$) of the covariance
matrix as a function of the temperature in the vicinity of
the critical point $c_2$ and $c_2'$ for $m^0_{\pi}=35$~MeV and $\mu=243$~MeV,$\mu=246$~MeV, $\mu=248$~MeV and $\mu=250$~MeV, respectively.}}
\label{s3}
\end{center}
\end{figure}
\begin{figure}[h!]
\begin{center}
\includegraphics[width=0.47\textwidth]{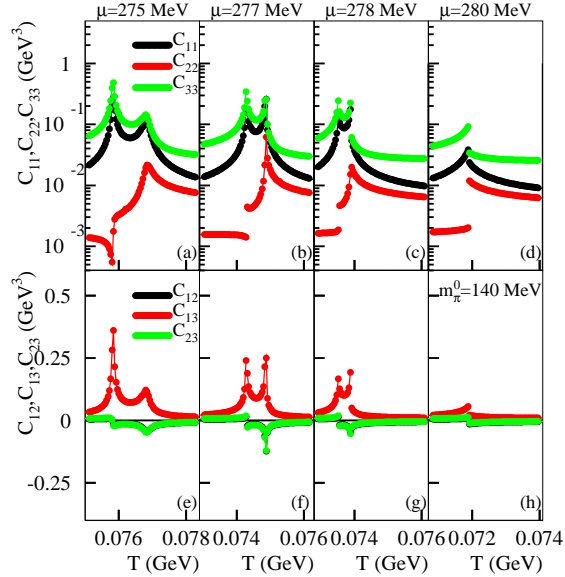}
\caption{\small{(Color online) Panels a, b, c and d: diagonal matrix elements of the covariance matrix 
in the vicinity of
the critical points $c_2$ and $c_2'$ as a function of the temperature for $m^0_{\pi}=140$~MeV 
and $\mu=275$~MeV,$\mu=277$~MeV, $\mu=278$~MeV and $\mu=280$~MeV, respectively.
Panels e, f, g and h: off-diagonal matrix elements of the covariance matrix 
in the vicinity of
the critical points $c_2$ and $c_2'$ as a function of the temperature for $m^0_{\pi}=140$~MeV 
and $\mu=275$~MeV,$\mu=277$~MeV, $\mu=278$~MeV and $\mu=280$~MeV, respectively.}}
\label{s4}
\end{center}
\end{figure}

 It's worth to mention that in our analysis, the correlations between $\rho_{\q}$ and $s$ are positive, whereas 
the correlations between $\rho_s$ and $s$, and $\rho_s$ and $\rho_{\q}$ are negative.
This is due to the fact that, at the transition, the thermal contribution to the scalar density drops from its maximum
value (in the broken phase) to a very small value (exactly zero in the chiral limit) in the symmetric phase 
(see ref.~\cite{Mocsy:2004ab}), whereas
both $s$ and $\rho_{\q}$ exhibit the opposite behavior going from smaller values (in the broken phase) to higher values 
(in the symmetric phase). Unlike $s$ and $\rho_{\q}$ the scalar density is, in fact, dominated by the zero-point 
contributions. With their inclusion one recover the physically expected behavior~\cite{Mocsy:2004ab}. One must then 
bear in mind
that the following results are relevant for the thermal part of the model. Our general conclusions, however, are 
in agreement with the results obtained with the NJL in sec.~\ref{secnjl} where the zero-point contributions are included.

\begin{figure}[h!]
\begin{center}
\includegraphics[width=0.47\textwidth]{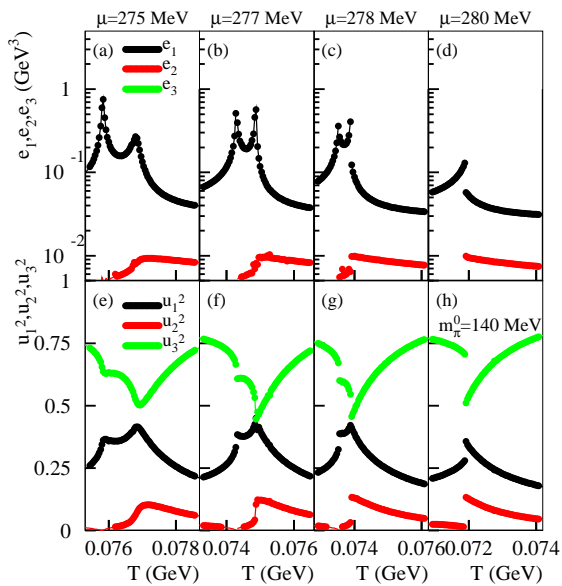}
\caption{\small{(Color online) Panels a,b,c and d: eigenvalues of the covariance matrix as a function of the temperature in 
the vicinity of the critical points $c_2$ and $c_2'$ for $m^0_{\pi}=140$~MeV and 
$\mu=275$~MeV, $\mu=277$~MeV, $\mu=278$~MeV and $\mu=280$~MeV, respectively.
Panels e,f,g and h: square components of the (normalized) eigenvector corresponding to the larger eigenvalue ($e_1$) of the covariance
matrix as a function of the temperature in the vicinity of
the critical point $c_2$ and $c_2'$ for $m^0_{\pi}=140$~MeV and $\mu=275$~MeV,$\mu=277$~MeV, $\mu=278$~MeV and
$\mu=280$~MeV, respectively.}}
\label{s5}
\end{center}
\end{figure}

We analyze two different values for the vacuum pion mass: $m^0_{\pi}=35$~MeV 
(figs.~\ref {s1},\ref {s2} and \ref {s3}) and $m^0_{\pi}=140$~MeV (figs.~\ref {s4} and \ref {s5}).
For clarity, it is convenient to chose a label for the various critical points. We will call $c_1$ the 
critical endpoint of the leftmost first order line (if there is any) that starts at $\mu=0$ and $T \sim 140$~MeV 
in the phase diagram. 
With $c_2$ and $c_2'$ we will refer to the two critical endpoints of the first order line 
(that ultimately splits in two) that starts at $T=0$ and $\mu\sim 300$~MeV, $c_2'$ being the rightmost 
of the two as in Fig.~\ref{f4}. As we have seen in the last section, for $m^0_{\pi}=10$~MeV there is only the critical point $c_2'$. 
We will not consider that case here, however.
We begin with $c_1$ for $m^0_{\pi}=35$~MeV. 
In Fig.~\ref {s1}, panels (a-b) and (c-d) we show
the three diagonal components of the covariance matrix ($\mathbf{C}_{11}$, $\mathbf{C}_{22}$, $\mathbf{C}_{33}$), and
the three off-diagonal components  ($\mathbf{C}_{12}$, $\mathbf{C}_{13}$, $\mathbf{C}_{23}$) for $\mu=70$~MeV and 
$\mu=86$~MeV. For $\mu=70$~MeV, the system undergoes a first order phase transition at $T=138.6$~MeV, 
whereas for $\mu=86$~MeV the transition is continuous (but still very sharp) and takes place at $T=137$~MeV. 
In between, one finds the unusual  critical point $c_1$. 
By looking at the elements of the covariance matrix for these two values of $\mu$ one sees 
that the dominant fluctuations are given by the entropy density and the scalar density. This is shown 
in Fig.~\ref {s1}a and \ref {s1}b, where the dominant diagonal terms in the vicinity of the transition 
are $\mathbf{C}_{22}=T\partial \rho_{s}/\partial \eta$, $\mathbf{C}_{33}=T\partial s/\partial T$.  
From the off-diagonal terms in Fig.~\ref {s1}c and \ref {s1}d one notices that the scalar and the entropy density 
are the most strongly anticorrelated (the magnitude of
$\mathbf{C}_{23}=T\partial \rho_{s}/\partial T$ is  much larger than the other 
off-diagonal coefficients).
This fact can be observed in a more direct way by looking at the eigenvalue $e_1$ and the corresponding 
eigenvector ${\bf u}$.
In Fig.~\ref {s1}e and \ref {s1}f the three eigenvalues of the covariance matrix are plotted for the same range of 
values of $T$ and $\mu$.
The largest eigenvalue 
$e_1$ is the only one showing a peak (it should actually diverge at the critical point).
The eigenvector ${\bf u}$ (see Eq.~(\ref {divdens})) gives us the three components of the critical density 
$\rho_{\bf u}=u_1\rho_{\q}+u_2\rho_{s}+u_3s$. As one can see, from Fig.~\ref {s1}g and \ref {s1}h, the 
critical density $\rho_{\bf u}$ is a mixture of almost solely the scalar and the entropy density 
(the first component $u_1$ does not appear in the plot as it is always very close to zero).  

As we have already discussed, in addition to this critical point the system exhibits 
two more critical points ($c_2$ and $c_2'$) 
in the high-$\mu$ region (see Fig.~\ref {f4}). 
The analysis of these two points is shown in Fig.~\ref {s2} and Fig.~\ref {s3} for $\mu=243,246,248,250$~MeV.
 For $\mu=243$~MeV (panels a and e in Fig.~\ref {s2} and \ref {s3}) the transition is continuous, but one 
can already clearly distinguish that the two peaks associated with the two critical points have already developed.
The first one on the left (the one corresponding to $c_2$) is very sharp, while the other (a little bit on the right,
corresponding to $c_2'$) looks still like a bump.
For $\mu=246$~MeV (panels b and f in Fig.~\ref {s2} and \ref {s3}) we have a first order and a continuous transition.
The first peak on the left is now a discontinuity, whereas the bump becomes sharper as the critical point $c_2'$ is
approached.
For $\mu=248$~MeV we have a double first order line (panels c and g in Fig.~\ref {s2} and \ref {s3}) and for 
$\mu=250$~MeV the two first order lines have merged together to form a single transition line. 

Looking at the diagonal terms of the covariance matrix in the upper panels of Fig.~\ref {s2} one immediately sees 
that the density $\rho_{\q}$ is now playing a relevant role and its fluctuations are of the same order of those of 
the scalar density. 
The off-diagonal terms (lower panels in Fig.~\ref {s2}) show a positive correlation between 
$\rho_{\q}$ and $s$ and a negative correlation between $\rho_s$ and $s$, and $\rho_s$ and $\rho_{\q}$.

The two critical points $c_2$ and $c_2'$, indeed, show different features. 
From the analysis of the diagonal matrix, one sees (lower panels in Fig.~\ref {s3}) that the critical density 
$\rho_{\bf u}$ is in both cases dominated by the entropy density. However, 
For the point $c_2$ (see Fig.~\ref{s3} panels a and e) the scalar component is rather small (Fig.~\ref{s3}e), 
and overcomes the net-quark number density component only 
at a temperature immediately higher than the one corresponding to the peak of the eigenvalue $e_1$ 
(compare to Fig.~\ref{s3}a). For the critical point $c_2'$, instead, the critical density 
$\rho_{\bf u}$ is $\sim 50\%$ entropy density and the remnant $50\%$ is almost equally split between 
net-quark and scalar density (see Fig.~\ref{s3}f).

The same analysis has been repeated for $m^0_{\pi}=140$~MeV in figs.~\ref {s4} and \ref {s5} in the vicinity 
of the two critical points $c_2$ and $c_2'$ (the point $c_1$ does not exist) 
for $\mu=275,277,278,280$~MeV. Once again, for $\mu=275$~MeV the transition is continuous,
whereas for $\mu=277$~MeV we have a first order and a continuous transition (corresponding to the rightmost spike), for 
$\mu=278$~MeV we have a double first order line and finally, for $\mu=280$~MeV the two first order lines have merged 
together to form a single one. By looking at the diagonal terms of the covariance matrix 
(upper panels in Fig.~\ref {s4}) one can see that now the fluctuations of both critical points are dominated by the 
net-quark density and the entropy density. In addition, one notices that in all the lower panels of Fig.~\ref {s4}, 
the correlation of the scalar density with the entropy density and the net-quark density are almost identical.
Even though in this case the scalar density plays only a minor role, 
it behaves in an even more visibly different way in the two critical points $c_2$ and $c_2'$. 
In Fig.~\ref {s4} (panels a and e) at 
$T=75.8$~MeV we are very close to the critical point $c_2$. 
In correspondence to this temperature, the scalar fluctuations
$\mathbf{C}_{22}$ are suddenly suppressed. At the same time, the correlations between  
$\rho_{s}$ and $\rho_{\q}$ ($\mathbf{C}_{12}$) 
and the correlations between $\rho_{s}$ and $s$ ($\mathbf{C}_{23}$) cross zero 
(this can be seen, with some difficulty, in Fig.~\ref {s4}e). 
Indeed, in correspondence to the leftmost critical point $c_2$, 
the scalar density seems to behave as a ``spectator'' of the critical phenomenon as it does not mix with the other
densities. 
This is not the case for the critical point $c_2'$. Even if smaller than the entropy and net-quark number density 
fluctuations, now the scalar density fluctuations show a peak and the correlations $\mathbf{C}_{12}$ 
and $\mathbf{C}_{23}$ are negative.

Increasing the mass of the pion the critical points $c_2$ and $c_2'$ move to 
the right of the phase digram towards higher values of the chemical potential and lower temperature. 
As a result, the fluctuations of the net-quark number density becomes more important. 
In contrast, the critical point $c_1$ (when it exists), 
has only a minor component of the net-quark number density, and the transition is dominated by the
entropy density and the scalar density, reflecting a more ``chiral'' behavior. This is in part due to the
fact that the critical point $c_1$ exists only when 
the vacuum pion mass is small, i.e. when the chiral symmetry is only slightly (explicitly) broken.

\section{The Nambu--Jona-Lasinio Model}
\label{secnjl}

Let us now consider the standard version of 
the two flavor NJL model which is described by~\cite{njl}

\begin{equation}
\mathcal{L}={\bar \psi}\left[i{\gamma }_{\mu }{\partial }^{\mu
  } -m_{c}%
  \right] \psi +G\left[ ({\bar \psi}\psi)^{2}+({\bar{\psi}} i\gamma
  _{5}{\tauv}\psi )^{2}\right] ,  
\label{njl2}
\end{equation}
where $\psi$ (a sum over flavors and color degrees of freedom is
implicit) represents a flavor isodoublet (u and d type of quarks)
$N_{c}$-plet quark fields while $\tauv$ are isospin Pauli
matrices. The Lagrangian density (\ref {njl2}) is invariant under
(global) $U(2)_{f}\times SU(N_{c})$ and, when $m_{c}=0$, the theory is
also invariant under chiral $U(2)_{L}\times U(2)_{R}$ groups. Note  that, as emphasized in 
Refs.~\cite{koch,buballa}, the  introduction of a vector
interaction term of the form $({\bar \psi} \gamma^\nu \psi)^2$ in
Eq. (\ref{njl2}) is also allowed by the chiral symmetry and that such a term can
become important at finite densities, generating a saturation mechanism
depending on the vector coupling strength that provides better matter
stability.
Regarding analytic nonperturbative evaluations, one can consider one
loop contributions dressed up by a fermionic propagator, whose
effective mass is determined in a self-consistent way. This
approximation is known under different names, e.g., Hartree,
large-$N_c$ or mean-field approximations (MFA).  To obtain the effective
potential (or Landau free energy density) for the quarks, $U_{eff}$, it is convenient to consider the bosonized version of the NJL,
which is easily obtained by introducing auxiliary fields ($\sigma ,{
  \piv}$) through a Hubbard-Stratonovich transformation.  Then, to introduce the auxiliary
bosonic fields and to render our results  more suitable to compare with  the
large-$N_{c}$ approximation it is convenient to use $G\rightarrow
\lambda /(2N_{c})$ and to formally treat $N_{c}$ as a large number,
which is set to the relevant value, $N_{c}=3$, at the end of the
evaluations. One then has

\begin{equation}
\mathcal{L}={\bar \psi}\left(i{\gamma }_{\mu }{\partial }^{\mu
  } -m_{c}\right) \psi -{\bar \psi}(\sigma +i\gamma
_{5}{\tauv \cdot }{\piv} )\psi -\frac{N_{c}}{2\lambda
}(\sigma ^{2}+{\piv}^{2}).
\label{njlboson}
\end{equation}
The Euler-Lagrangian equations show that $\sigma=-(\lambda/N_c) {\bar \psi} \psi=-2G {\bar \psi} \psi$ and 
$\piv=-(\lambda/N_c) {\bar \psi} i\gamma_{5}{\tauv}\psi=-2G {\bar \psi}i\gamma_{5}{\tauv} \psi$.

\subsection{Optimized Perturbation Theory for the NJL model}

\begin{figure}[tbh]
\vspace{0.5cm} 
\epsfig{figure=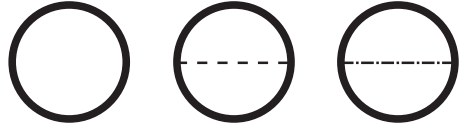,angle=0,width=6cm}
\caption{Diagrams contributing to $U_{eff}\left(
  {\hat{\protect\eta}} \right) $ to order $\protect\delta$. The thick
  continuous fermionic lines represent $\ {\hat{\protect\eta}}= m_c +\eta -\delta( \eta-\sigma)$
  dependent terms which must be further expanded. The dashed lines
  represent the $\protect\sigma $ propagator and the $\protect\pi $
  propagator is represented by dashed-doted line. The first
  contributes with $1/N_{c}^{0}$, the second and third diagrams (of
  order $\protect\delta $) contribute with $1/N_{c}$.}
\label{effpot}
\end{figure}

The basic idea of the OPT method is to deform the original Lagrangian density by adding a quadratic term like  $(1-\delta)\eta \bar{\psi} \psi$ to the original Lagrangian density as well as multiplying all coupling constants by $\delta$. The new parameter $\delta$ is just a bookkeeping label and $\eta$ represents an {\it arbitrary} mass parameter \footnote {Note that although we have kept the same notation, the  $\eta$ used here has as completely different role than the one used previously in the L$\sigma$M.}. Perturbative calculations are then performed in powers of the dummy parameter $\delta$ which is
formally treated as small and set to the original value, $\delta=1$,  at the end\footnote {Recall that within the large-$N_c$ on performs an expansion in powers of $1/N_c$ where $N_c$ is formally treated as large but set to the original value ($N_c=3$ in our case) at the end.}. Therefore, the fermionic propagators is dressed by  $\eta$ which may also be viewed as an infra red regulator in the case of massless theories. After a physical quantity, such as $U_{eff}$, is evaluated to the order-$k$ and $\delta$ set to the unity a residual $\eta$ dependence  remains. Then,  optimal  non perturbative results can be obtained by requiring that $U_{eff}^{(k)}(\eta)$ be evaluated where it is less sensitive to variations of the arbitrary mass parameter. This requirement translates into the criterion known as the Principle of Minimal 
Sensitivity (PMS)~\cite {pms}
\begin{equation}
\left. \frac{d{ U_{eff}}^{(k)}(\eta)}{d\eta }\right\vert _{\bar{\eta},\delta
  =1}=0\;.
\label{pms12}
\end{equation}
In general, the solution to this equation implies in self consistent relations  generating a non perturbative $G$ dependence.
In most cases non perturbative   $1/N_c$ corrections appear  already at the first non trivial order while the MFA results can be recovered at any time simply by considering $N_c \to \infty$. Finally, note that the OPT has the same
spirit as the Hartree and Hartree-Fock approximation in which one also adds and subtracts a mass term. However, within these
two traditional approximations the topology of the dressing is fixed from the start: direct (tadpole) terms for Hartree and
direct plus exchange terms for Hartree-Fock. On the other hand, within the OPT $\bar \eta$ acquires characteristics which change order
by order progressively incorporating direct, exchange, vertex corrections, etc. effects.  
To implement the OPT within the NJL model one follows the prescription
used in Refs.~\cite{prdgn2d,prdgn3d, prc} by first interpolating the
original four-fermion version. Then, in terms of the auxiliary fields,
the deformed Lagrangian density  becomes 

\begin{equation}
\mathcal{L}=\bar{\psi}\left[ i{\gamma }_{\mu }{\partial }^{\mu
  }-m_{c}-\delta \left( \sigma +i{\gamma }_{5}{\tauv\cdot
    \piv} \right) -\eta \left( 1-\delta \right) \right] {\psi
}-\delta \frac{N_{c} }{2\lambda }\left( \sigma
^{2}+{\piv}^{2}\right) \,,  
\label{delta12}
\end{equation}
which shows that the Yukawa vertices have weight $\delta$ while the meson ``propagators'' are proportional to $1/\delta$. One is then ready to perform  a {\it perturbative} evaluation of Landau's  free energy in powers of $\delta$.  In the $\sigma$ direction, the first non trivial order the relevant contributions are represented in 
Fig.~\ref  {effpot}. Then, at finite temperature and finite chemical potential, the order-$\delta$ result for the NJL free energy density is 
(see Ref.~\cite {prc} for a more detailed discussion)

\begin{eqnarray}
{U_{eff}}(\sigma)& =&\frac{\sigma^{2}}{4 G}-2N_{\rm f}N_c I_1 (\mu,T) +2
\delta N_{\rm f}N_c(\eta+m_c) \left( \eta-\sigma \right)
I_2(\mu,T) \notag\\  && + 4\delta G N_{\rm f}N_{c} \:I_3^{2}(\mu,T) -
2\delta G  N_{\rm f}N_{c}\,(\eta+m_c)^2 I^2_2(\mu,T)  \,\,\,,
\label{landauenergy}
\end{eqnarray}
where we have replaced $\lambda \to 2 G N_c$.  In the above equation
we have defined, for convenience, the following basic relevant
integrals:

\begin{equation}
I_1(\mu,T) = \int \frac{d^{3}p}{\left( 2\pi \right) ^{3}}\left\{ \varepsilon +
T\ln \left[ 1+e^{-\left( \varepsilon+\mu \right) /T}\right] +T\ln \left[
  1+e^{-\left( \varepsilon-\mu \right) /T}\right]  \right\} \;,
\end{equation}

\begin{equation}
I_2(\mu,T) = \int \frac{d^{3}p}{\left( 2\pi \right)^{3}}
\frac{1}{\varepsilon}\left[ 1-\frac{1}{e^{\left( \varepsilon+\mu \right)
      /T}+1}-\frac{1}{e^{\left( \varepsilon-\mu \right) /T}+1}\right]\;,
\end{equation}
and

\begin{equation}
 I_3(\mu,T) =\int \frac{d^{3}p}{\left( 2\pi \right)^{3}} \left[
   \frac{1}{e^{\left( \varepsilon-\mu \right) /T}+1}-\frac{1}{e^{\left(
       \varepsilon+\mu \right) /T}+1}\right]\;,
\label{basicint}
\end{equation}
where $\varepsilon^{2}={\bf p}^2+(\eta+m_c)^{2}$. Notice also that $I_3$ only survives at $\mu \ne 0$. Here, we
impose a sharp non-covariant cut-off, $\Lambda$, only for the vacuum term, since the finite temperature has
a natural cut-off in itself specified by the temperature. This choice of
regularization, which allows for the Stefan-Boltzmann limit  to be reproduced
at high temperatures, is sometimes preferred  in the
literature~\cite{Fukushima:2008is, prc, pedro}. Moreover, in the present application it assures that the temperature integrals  appearing in  
the L$\sigma$M and in the NJL are integrated over the same momentum range.  Also, as will be further 
discussed, this regularization choice appears to be a crucial condition in order for 
the NJL model to reproduce the phase diagram with two critical points.

The divergent integrals
occurring at $T=0$ and $\mu=0$ are

\begin{eqnarray}
I_1(0,0) &=& \int \frac{d^{3}p}{\left( 2\pi \right) ^{3}} \varepsilon
\nonumber \\ &=& \frac{1}{32\pi ^{2}} \left\{ (\eta+m_c)^{4} \ln
\left[ \frac{ \left( \Lambda + \sqrt{ \Lambda^{2}+(\eta+m_c)^{2}}
    \right)^{2}} {(\eta+m_c)^{2}}\right] -
2\sqrt{\Lambda^{2}+(\eta+m_c)^{2}}\left[2\Lambda^{3}+ \Lambda
  (\eta +m_c)^{2}\right] \frac{}{}\right \}\,,
\label{I1div}
\end{eqnarray}
and

\begin{equation}
I_2(0,0) = \int \frac{ d^{3}p}{\left( 2\pi \right)^{3}}
\frac{1}{\varepsilon}=\frac{1}{4\pi^2} \left\{ \Lambda
\sqrt{\Lambda^{2}+(\eta+m_c)^{2}} -\frac{(\eta+m_c)^{2}}{2} \ln
\left[ \frac{\left[ \Lambda+\sqrt{\Lambda^{2}+
        (\eta+m_c)^{2}}\right]^{2}}{(\eta+m_c)^{2}}\right]
\right\}\;.
\label{I2div}
\end{equation}
If needed, the  $T \to 0$ limit of those integrals can be readily obtained (see Ref.~\cite {prc}).
Then, by applying the PMS relation to $U_{eff}$ one gets

\begin{equation}
 \left\{ \left[\eta - \sigma - 2 (\eta+m_c) G \, I_2\right] \left[ 1 +
   (\eta+m_c) \frac {d}{d \eta} \right] I_2 + 4 G \, I_3 \frac{d}{d \eta} I_3
 \right\}_{\eta = {\bar \eta}}=0 \,.
\label{simplePMS}
\end{equation}
Notice that if one ignores the terms proportional to $G$ (which are of order $1/N_c$)  the optimal result is simply ${\bar \eta=\sigma}$. In this situation,   Eq. (\ref  {landauenergy}) shows that the MFA result is exactly reproduced. 
Since we are mainly interested in the thermodynamics, one basic quantity of
interest is the thermodynamical potential, $\Omega$, whose relation to the
the free energy is given by $\Omega= {U_{eff}}({\bar \sigma})$. The order
parameter, ${\bar \sigma}$ is determined from the gap equation generated by
minimizing ${U_{eff}}$ with respect to $\sigma$. {}From Eq. (\ref {landauenergy}) we obtain  \cite {prc}

\begin{equation}
 {\bar \sigma} = 4 G N_{\rm f} N_c (\eta+m_c) I_2\;.
\label{simpleGAP}
\end{equation}
In order to further discuss the relation of some of our results with those obtained by  Fukushima~\cite{Fukushima:2008is} it is interesting to note, at this stage, that 
 since ${\bar \sigma}= \langle \sigma \rangle= - 2 G \langle {\bar \psi} \psi \rangle = -2G \rho_s$ one can write the optimum thermodynamical potential as
\begin{equation}
 \Omega= G \rho_s^2 +\Omega_{\rm MFAL}(\bar \eta)- ({\bar \eta} + 2G\rho_s)\rho_s + \frac{G}{N_c N_f}\left ( \rho_{\q}^2-\frac{\rho_s^2}{2}\right) \,\,,
\label{omegarhos}
\end{equation}
where $\Omega_{\rm MFAL}(\bar \eta)$ has the mathematical structure of a MFA like thermodynamical potential whose effective mass given by $\bar \eta$. Since $\Omega =-P$ one sees that the OPT introduces a correction like $-G/(N_c N_f) \rho_{\q}^2$ which, although suppressed by $1/N_c$,  is of the same form as the one considered by the MFA when a vector interaction term like $-G_V {\bar \psi} V\gamma_\mu V^\mu \psi$, as
proposed in Ref.~\cite{koch}, is added to the original NJL Lagrangian density. 
When  $N_c \to \infty$, ${\bar \eta} = -2G \rho_s$ and the MFA result (for the standard NJL model) with no $\rho_{\q}$ dependence is recovered.

\subsection{Non Standard Parametrization and the Appearance of Two Critical Points}
\label{nonstandard}
Usually, the three NJL parameters, $G$, $m_c$ and $\Lambda$ are fixed by fitting the 
pion decay constant, $f_\pi= 92.4 \, {\rm MeV}$, the quark condensate  $190 {\rm MeV} \lsim -\langle {\bar \psi} \psi \rangle^{1/3} \lsim 260 \, {\rm MeV}$  as well as the pion mass, $m_\pi = 135 \, {\rm MeV}$ 
leading, in the MFA, to values such as  $\Lambda = 664.3 \, {\rm MeV}$, $G \Lambda^2 =2.06$ and 
$m_c=5 \, {\rm MeV}$~\cite {buballa}. With these values one obtains satisfactory predictions for 
the quark vacuum effective mass, $m_q^0$, and for the quark condensate, $\langle {\bar \psi} \psi \rangle$ given 
by $m_q^0= 300 \,{\rm MeV}$ and $-\langle {\bar \psi} \psi \rangle^{1/3}= 250.8 \, {\rm MeV}$.
Although in general the non covariant cut-off lies within the range $500-700 \, {\rm MeV}$  while 
$m_c \sim 5 \, {\rm MeV}$ the coupling can be further increased producing much higher values for $m_q^0$ without affecting too much the $G$ independent quark condensate. 
For example, another 
MFA set also presented in Ref.~\cite {buballa} is given by $\Lambda = 568.6\, {\rm MeV}$, 
$G \Lambda^2 =3.17 $ and $m_c=5.1 \, {\rm MeV}$ predicting $m_q^0= 600 \, {\rm MeV} > \Lambda$  and 
$\langle {\bar \psi} \psi \rangle= -247.5 \, {\rm MeV}$.
As one can see the value of $m_q^0$ doubles while that of the observable $\langle {\bar \psi} \psi \rangle$ 
remains within the bounds $190 {\rm MeV} \lsim -\langle {\bar \psi} \psi \rangle^{1/3} \lsim 260 \, {\rm MeV}$ set by sum rules ~\cite {SR} and the value $-\langle {\bar \psi} \psi \rangle^{1/3} \simeq 231 \, {\rm MeV}$ which corresponds to lattice estimates ~\cite{lat}. Regarding 
applications at finite temperature and density one sees that also the values of the critical temperature  (at $\mu=0$) and of the critical chemical potential  (at $T=0$) increase with $G$. In general,  the size of  the first order transition line, 
which originates at high $\mu (\sim m_q^0)$ and $T=0$, increases with $G$ approaching the $T$ axis for 
very high coupling strengths. 
This observation, together with the L$\sigma$M results that the appearance of two critical points becomes 
possible for $m_\pi < 50 \, {\rm MeV}$ gives us the hint to use the OPT at high $G$ and small $m_c$ 
(since $m_\pi \propto m_c$) while setting $\Lambda$ to usual values.
For consistency we must use the OPT two loop relations for $f_\pi$ and $m_\pi$ recently found in 
Ref.~\cite {prc} and which predict some deviations from the MFA result for the 
Gell--Mann-Oakes-Renner relation. Taking  $\Lambda = 590 \,  {\rm MeV}$, and  $G \Lambda^2=3.7$ 
with $m_c=4.5 \, {\rm MeV}$ one obtains the very reasonable values $f_\pi \simeq 92\, {\rm MeV}$, 
$m_\pi \simeq 135 \, {\rm MeV}$, and $-\langle {\bar \psi} \psi \rangle^{1/3} \simeq 264 \, {\rm MeV}$. 
However,  now $G$ has an extremely high value which is reflected in the vacuum effective quark 
mass value , $m_q^0 \simeq 787 \, {\rm MeV}$.
Next, one can keep those values of $\Lambda$ and $G$  decreasing $m_c$  so as to make contact 
with the L$\sigma$M results. For example, $m_c=0.1, 0.28,0.55 \, {\rm MeV}$ lead to 
$m_\pi=20, 35, 50 \, {\rm MeV}$
while $f_\pi$ and $\langle {\bar \psi} \psi \rangle^{1/3}$ remain very stable. With this aim let 
us analyze the phase transition pattern at high $T$ and $\mu =0$ in search for a first order transition 
taking $m_c=0.1\, {\rm MeV}$ (with $\Lambda=590 \, {\rm MeV}$ and $G\Lambda^2=3.7$). Within the OPT this 
choice predicts $f_\pi \simeq 92.2 \, {\rm MeV}$, $m_\pi \simeq 20\, {\rm MeV}$ and 
$-\langle {\bar \psi} \psi \rangle^{1/3} \simeq 263 \, {\rm MeV}$ with the expected  
high quark mass value, $m_q^0=781 \, {\rm MeV}$. In principle, one could object to the fact that $m_q^0 > \Lambda$ (apart from having a numerical value much higher than 1/3 of the baryonic mass). However, as emphasized in the introduction, the  goal of our  
NJL investigation has a more qualitative character and, at the same time, it will be shown that the relevant temperature and chemical potential for our purposes of finding evidence for a second critical 
point fall well below $\Lambda$. Finally, note that with our parameter choice the large value of $G$ mostly affects  $m_q^0$ while physical  observables such as $f_\pi$, $m_\pi$, and $\langle {\bar \psi} \psi \rangle^{1/3}$ remain realistic.

\begin{figure}[tbh]
\vspace{0.5cm} 
\epsfig{figure=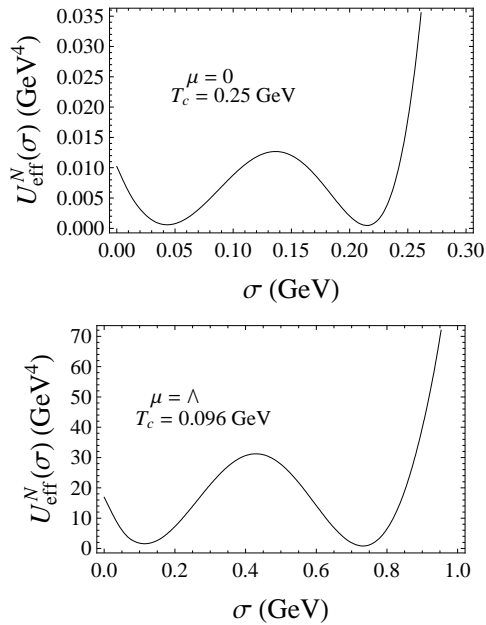,angle=0,width=7cm}
\caption{Normalized free energy, ${U_{eff}}^N (\sigma)= [{U_{eff}}(\sigma)- {U_{eff}}({\bar \sigma})]\times 10^{-7}$, as a function of $\sigma$ for ($\mu=0 \, {\rm MeV}$,$T_c=0.25 \, {\rm GeV}$) (top panel) and 
($\mu= \Lambda$,$T_c=0.096 \, {\rm GeV}$) (bottom panel) showing first order phase transition. }
\label{freenergy}
\end{figure}

With these unusual parameter values one indeed obtains a first order transition at $\mu=0$ and $T=0.25 \, {\rm GeV}$ 
as shown by the top panel of Fig.~\ref  {freenergy} while the usual first order transition is also 
observed at $T=0$ and $\mu_c \simeq 680 \, {\rm MeV} < m_q^0$ starting a line of first order transitions as 
shown by the bottom panel of  Fig.~\ref  {freenergy} which shows the degenerate minima at $\mu=\Lambda$ 
and $T=0.096 \, {\rm GeV}$. 

\begin{figure}[tbh]
\vspace{0.5cm} 
\epsfig{figure=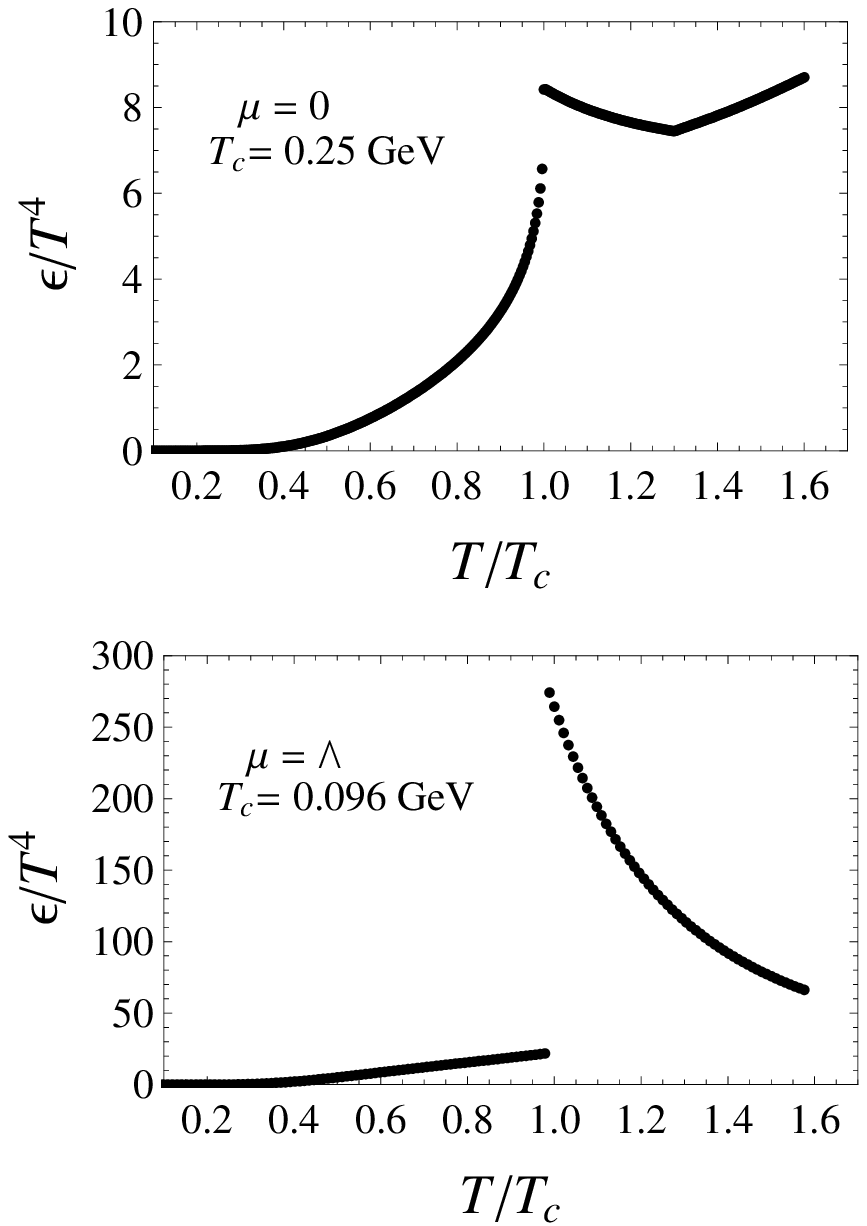,angle=0,width=7cm}
\caption{$\epsilon/T^4$ versus $T/T_c$  for $\mu=0, T_c=250 \,{\rm MeV}$ (top panel) and $\mu=\Lambda=590 \,{\rm MeV}, T_c=96 \, {\rm MeV}$ (bottom panel). The associated latent heat is $\Delta \epsilon= 0.3 \times 10^{-2} \, {\rm GeV}^4$ for the top panel and $\Delta \epsilon= 1.72 \times 10^{-2} \, {\rm GeV}^4$ 
 for the bottom panel. In both cases $G \Lambda^2=3.7$ and $m_c=0.1$.}
\label{latheat}
\end{figure}

The  differences in latent heat, $\Delta \epsilon$, between the two transitions can be further observed by looking at 
Fig.~\ref  {latheat} which shows the normalized $\epsilon/T^4$ versus the dimensionless ration $T/T_c$. The 
values are $\Delta \epsilon = 0.3 \times 10^{-2} \, {\rm GeV}^4$ for  $\mu=0$ and $T=250 \, {\rm MeV}$  and $\Delta \epsilon= 1.72 \times 10^{-2} \, {\rm GeV}^4$ for $\mu=\Lambda$ and  $T=96 \, {\rm MeV}$ (and of course higher for $T=0,\mu=680 \, {\rm MeV}$ but we shall refrain from numerically exploring the $\mu > \Lambda$ region).

\begin{figure}[tbh]
\vspace{0.5cm} 
\epsfig{figure=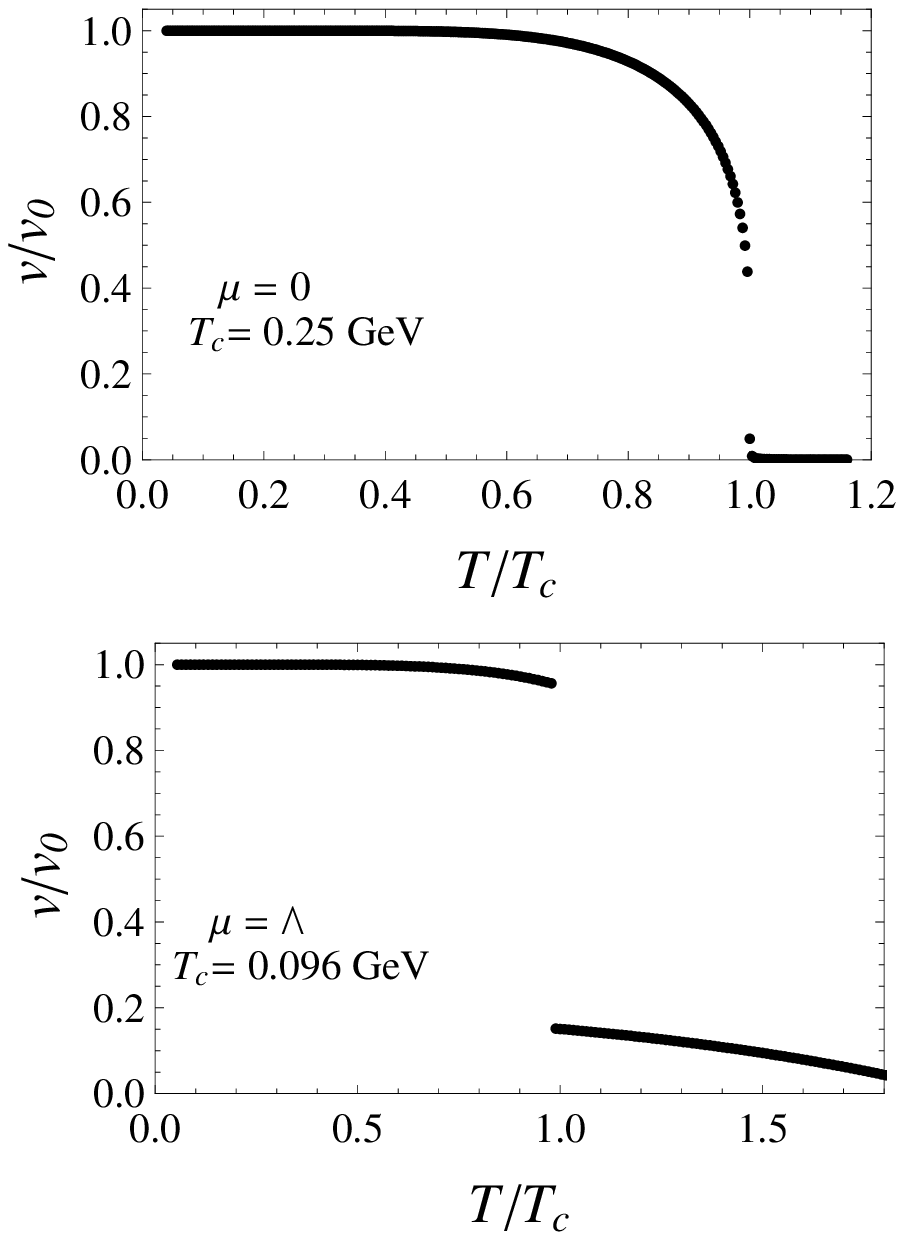,angle=0,width=7cm}
\caption{The dimensionless quark condensate ratio $v/v_0$ versus $T/T_c$  for $\mu=0, T_c=250 \, {\rm MeV}$ (top panel) and $\mu=\Lambda=590 \, {\rm MeV}, T_c=96 \, {\rm MeV}$ (bottom panel). In both cases $G \Lambda^2=3.7$ and $m_c=0.1$. }
\label{orparam}
\end{figure}

\begin{figure}[tbh]
\vspace{0.5cm} 
\epsfig{figure=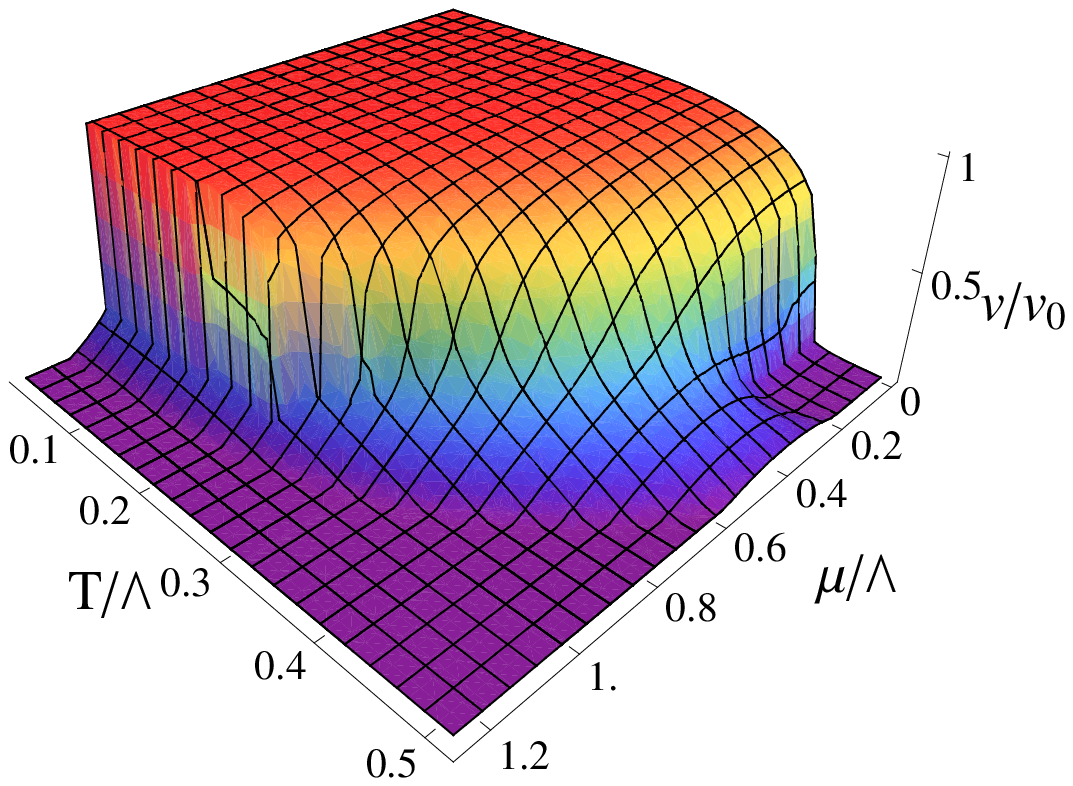,angle=0,width=7cm}
\caption{The dimensionless quark condensate ratio $v/v_0$ as a function of $T/\Lambda$ and $\mu/\Lambda$}
\label{PD3d}
\end{figure}

The situation can be also  observed by analyzing the thermal behavior of the order parameter represented by the quark condensate, $v=\langle {\bar \psi \psi} \rangle$, which is given in 
Fig.~\ref  {orparam} for $\mu=0$ and $\mu=\Lambda$. This figure clearly indicates that the first order happening at high $T$ and vanishing $\mu=0$ is much softer than the one happening in the reversed situation of small $T$ and high $\mu$, a fact which is well illustrated by the three dimensional plot of 
Fig.~\ref  {PD3d} which already shows that these two first order transition lines are indeed separated by a cross over region at intermediate $T$ and $\mu$. 

\begin{figure}[tbh]
\vspace{0.5cm} 
\epsfig{figure=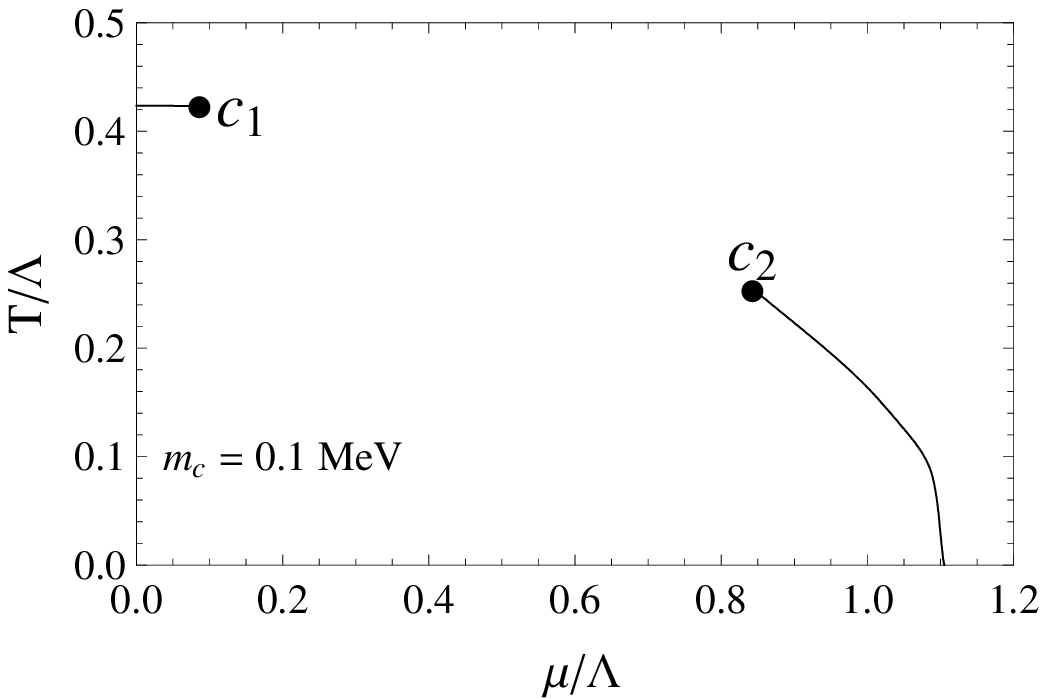,angle=0,width=7cm}
\caption{Phase diagram in the $T-\mu$ plane for $m_c=0.1 {\rm MeV}$ ($m_\pi \simeq 20 \, {\rm MeV}$) and corresponds to 
the choice $G \Lambda^2=3.7$. }
\label{PhDiaNJL}
\end{figure}

This fact can be more clearly appreciated  by projecting the first order lines of Fig.~ \ref  {PD3d} on the $T-\mu$ 
plane as in Fig.~\ref  {PhDiaNJL}. This phase diagram is qualitatively very similar to the one obtained in 
Fig.~\ref{f3} for the L$\sigma$M (as mentioned above our value $m_c=0.1$ leads to $m_\pi=20 \, {\rm MeV}$) and constitutes 
our first important result concerning the NJL model. Namely, that by going beyond the MFA and choosing the 
parameters so as to reproduce small pion masses one may obtain a
second critical end point in the phase diagram.  In addition, the
choice of regularization procedure appears to be crucial for the
appearance of $c_1$ in the NJL-model. If one  uses a cut off also in the $T$ dependent integrals the critical point $c_1$ does not emerge. However, some authors (see Ref.\cite {pedro}) have recognized that  a three dimensional cut off is only needed at zero temperature so that the presence of high momentum quarks in the $T$ dependent Feynman loops is required to ensure that the entropy density scales as $T^3$ at high temperature.  In contrast, by being renormalizable the \lsm always allows for high momentum quarks in the $T$ dependent loops. Therefore, with  our regularization choice the temperature dependent integrals within the NJL are treated in the same footing as their \lsm counterparts. A comprehensive discussion about how parameters and regularization affect NJL has recently been carried out  by Costa et al. \cite {pedro}.

For our parameter values 
the location of $c_1$ happens at the point ($T_{c_1}= 249.75 \, {\rm MeV}, \mu_{c_1}=51.16 \, {\rm MeV}$) and of $c_2$ at the point
($T_{c_2}= 149.78 \, {\rm MeV}, \mu_{c_2}=497.55 \, {\rm MeV}$). From the quantitative point of view these values are certainly high as one would expect from the fact that $m_q^0$ is also high. Nevertheless, the phase diagram resembles the one obtained with \lsm and which displays more reasonable numerical values. Note also how the first order line associated with $c_1$ is almost parallel to the $\mu$ axis.

\begin{figure}[tbh]
\vspace{0.5cm} 
\epsfig{figure=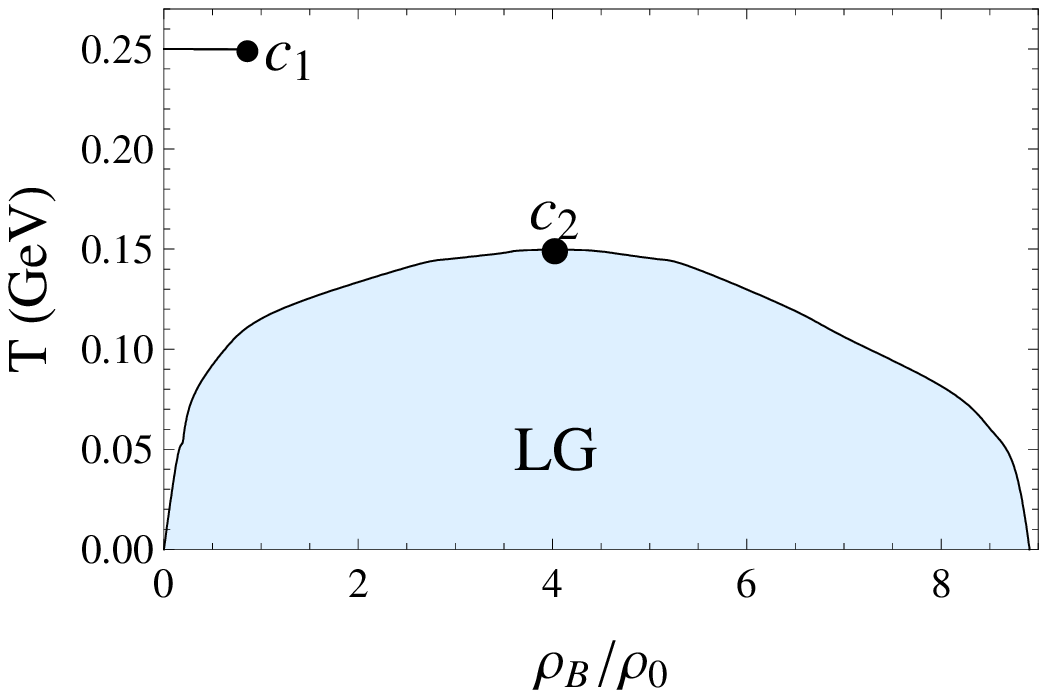,angle=0,width=7cm}

\epsfig{figure=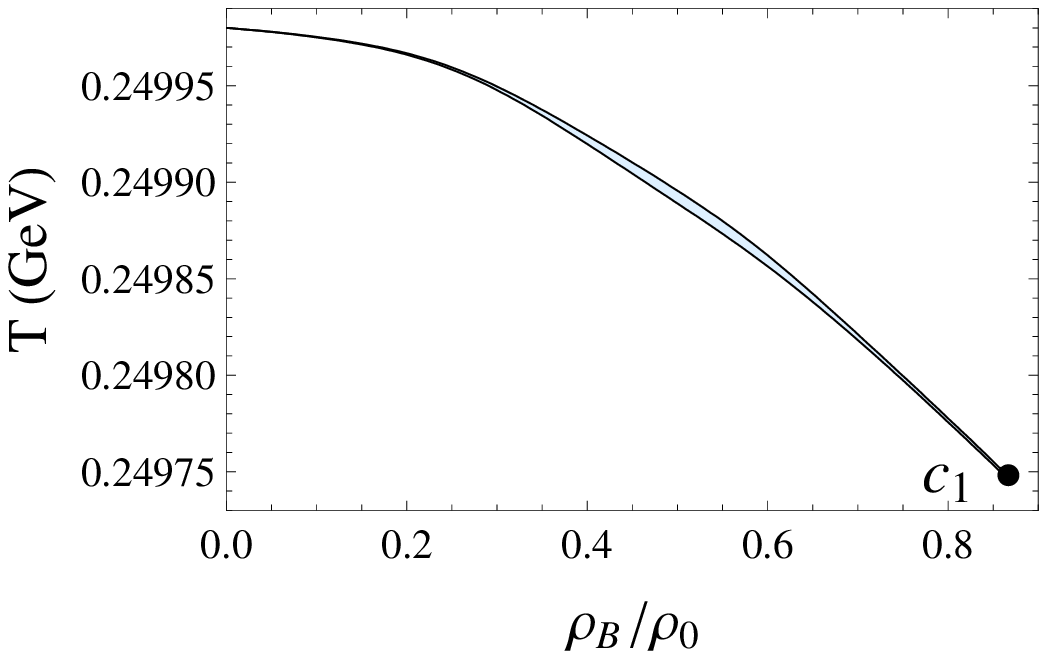,angle=0,width=7cm}
\caption{Phase coexistence diagram in the $T-\rho_B/\rho_0$ plane, where $ \rho_B = \rho/3$ is the baryonic density and $\rho_0=0.17 \, {\rm fm}^{-3}$ is the nuclear matter density. Here, $m_c=0.1 {\rm MeV}$ predicting  $m_\pi \simeq 20 \, {\rm MeV}$ with the choice $G \Lambda^2=3.7$. The two dark regions denote a mixed phase with LG denoting the liquid-gas type. The bottom panel is an expanded view of phase coexistence region associated with the high-$T$ first order transition line. }
\label{coeTrho}
\end{figure}

The next step is to understand the physical nature of both critical first order lines. Our study of the susceptibilities and densities in the L$\sigma$M has revealed that the high $T$ small $\mu$ line terminating at $c_1$ has a more ``chiral'' character   while the small $T$ and high $\mu$ line terminating at $c_2$ has a more hydrodynamical character.  One can then map the $T-\mu$ points to form a phase coexistence diagram such as the $T$ versus $\rho_B/\rho_0$ shown in 
Fig.~\ref  {coeTrho} where $\rho_0=0.17 \, {\rm fm}^{-3}$ is the
normal nuclear density. The top figure shows that the coexistence
region associated with the traditional $c_2$ point goes  from $T=0$
to $T=149.78 \, {\rm MeV}$ covering very high baryonic densities. The
unusual region, on the other hand, is very tiny going from $\rho_B=0$
to $\rho_B=0.85 \rho_0$  being restricted to a very narrow temperature
range as shown by the bottom panel. The critical points $c_1$ and
$c_2$ are located at $\rho_B=0.95 \rho_0$ and $\rho_B=4 \rho_0$ respectively. 

\begin{figure}[tbh]
\vspace{0.5cm} 
\epsfig{figure=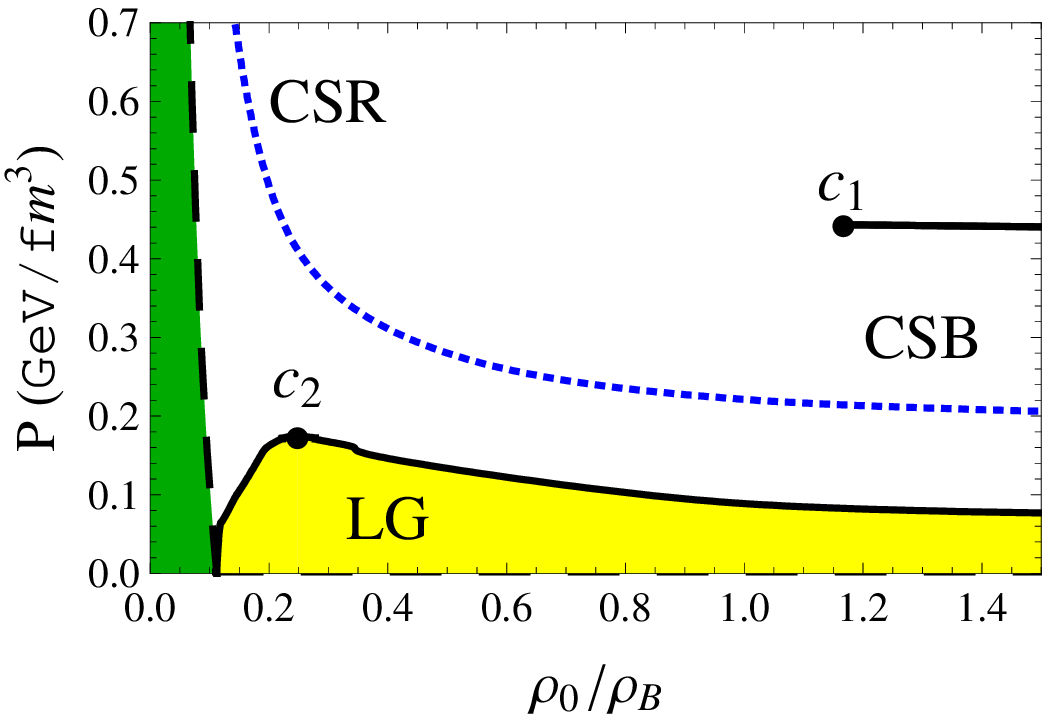,angle=0,width=7cm}

\caption{Phase diagram in the $P-\rho_0/\rho_B$ plane, where $ \rho_B = \rho/3$ is the baryonic density and $\rho_0=0.17 \, {\rm fm}^{-3}$ is the nuclear matter density. Here $m_c=0$ (chiral limit) and CSB represents the broken phase (``gas'') while CSR represents symmetric phase (``liquid'').  The long dashed line  is the $T=0$ isothermal and the dark region to its left is not accessible. The short dashed line represents the  $T= 220 \, {\rm MeV}$ isothermal corresponding to a cross over temperature.  The large mixed phase corresponding to the usual first order transition line of the liquid-gas (LG) type is labeled, the dot marks $c_2$.
 The thick continuous line at $P \simeq 0.45 \, {\rm GeV/fm^3}$ represents the region associated with the unusual line and the dot marks $c_1$. }
\label{PhDiaPV}
\end{figure}

At $\rho_B=0$ the  small coexistence region terminates into a point located at $T=250 \, {\rm MeV}$. Both coexistence regions have very different shapes, especially near the critical points, from which one may expect that the associated critical exponent $\beta$, to be evaluated later, acquires different values in each situation.
Let us now map the $T-\mu$ results into the $P-\rho_0/\rho_B$ plane as shown by Fig.~\ref  {PhDiaPV} which, again, clearly indicates that the region associated with $c_2$ in fact looks like the mixed liquid-gas phase appearing in $P-V$ type of phase diagram for a van der Waals fluid (which does not contain an analogous of the mixed phase associated with $c_1$). 

\subsection{Thermodynamical Quantities Near the Critical Points}

\begin{figure}[tbh]
\vspace{0.5cm} 
\epsfig{figure=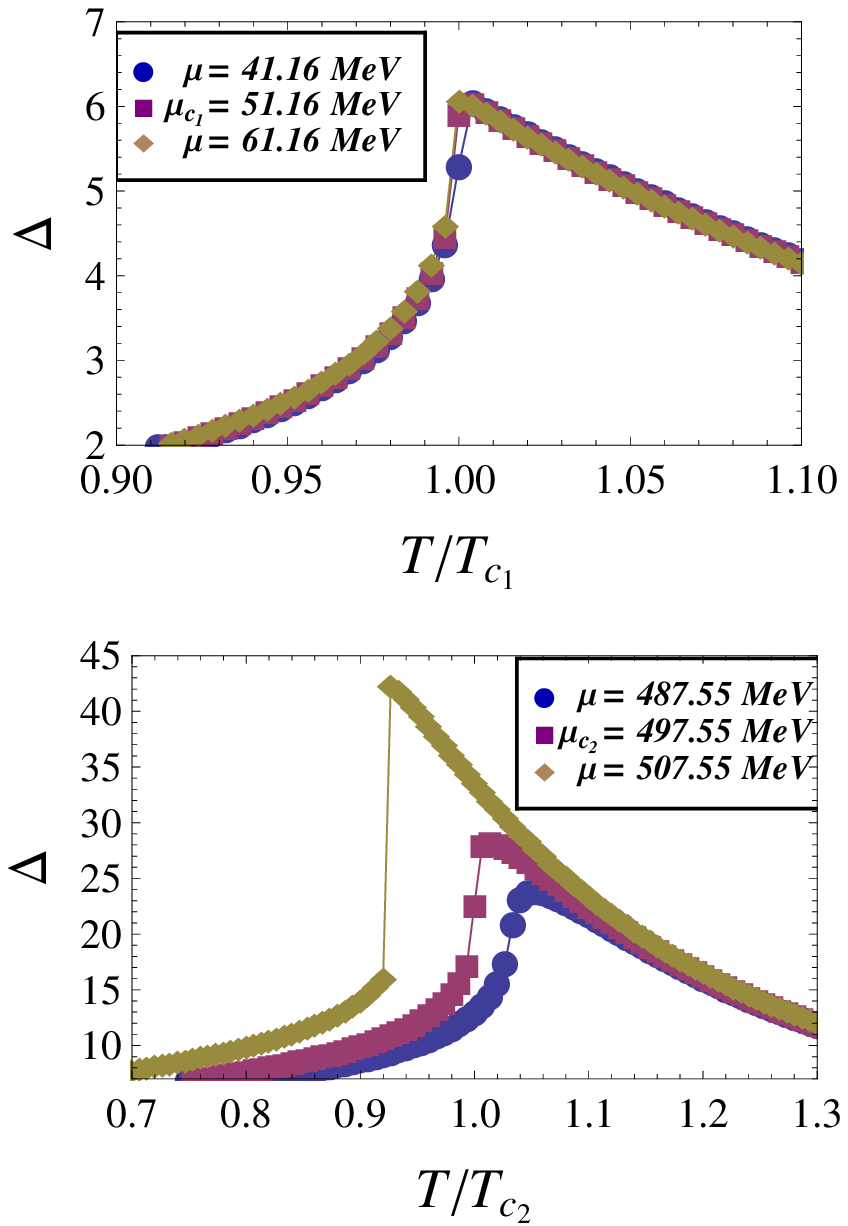,angle=0,width=7cm}
\caption{The interaction measure, $\Delta$, for the critical point $c_1$ (top panel) and for  
$c_2$ (bottom panel). Chemical potentials above and below the critical points values are also shown for  reference.In both cases $G \Lambda^2=3.7$ and $m_c=0.1$.}
\label{Delta}
\end{figure}

The OPT result for $\Omega=-P$ allows us to obtain $\epsilon = -P+Ts+\mu \rho_{\q}$ for the NJL with the 
inclusion of finite $1/N_c$ corrections which makes it possible to analyze the critical behavior  near each one of the two critical points. 
With this aim we will numerically evaluate  the following thermodynamical quantities: the 
interaction measure  (or trace anomaly, $\Delta$), the equation of state parameter ($w=P/\epsilon$), 
the bulk viscosity over entropy density ($\zeta/s$), the quark number and chiral susceptibilities ($\chi_{\q}$ and $\chi_m$) as well as some critical exponents.  Let us start by considering the interaction measure 
\begin{equation}
\Delta = \frac{(\epsilon - 3P)}{T^4}\;,
\label{Deltaeq}
\end{equation}
which is plotted in Fig. \ref {Delta} for regions near $c_1$ and $c_2$. This quantity is expected to peak near a phase transition or cross over and can be helpful in locating the critical line. Fig. \ref{Delta} indicates that the rise near $c_1$ looks more uniform around $c_2$ and happens for temperatures near $T_{c_1}$ which is not surprising since there is little variation in $T$ along the associated first order line as already emphasized. Although the peaks associated with $c_2$ look more pronounced one has to recall that $\Delta$ is normalized by $1/T^4$ and that the $c_2$ region is associated with lower temperatures.

\begin{figure}[tbh]
\vspace{0.5cm} 
\epsfig{figure=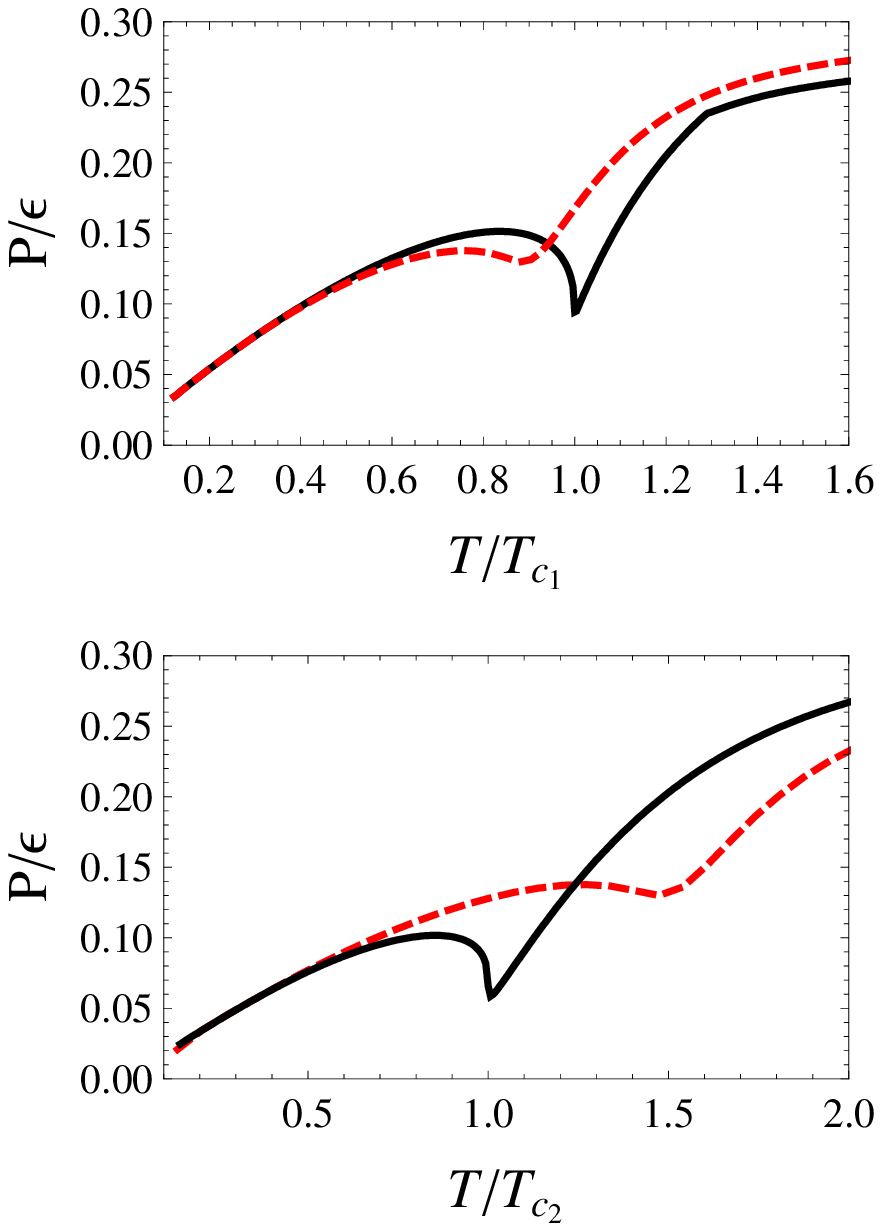,angle=0,width=7cm}
\caption{The EoS parameter $w=P/\epsilon$ versus the temperature  for the critical point $c_1$ (continuous line, top panel) 
and for $c_2$ (continuous line, bottom panel). In both cases $G \Lambda^2=3.7$ and $m_c=0.1$.  For ref reference we show the EoS parameter for $\mu> \mu_{c_1}$ (dashed line, top panel) and $\mu< \mu_{c_2}$ (dashed line, bottom panel). Here, $G \Lambda^2=3.7$ and $m_c=0.1$.}
\label{w}
\end{figure}

Next, let us investigate the EoS parameter, $w$, as represented by Fig. \ref {w} for the two critical points. In both situations one observes a downward cusp at the critical temperatures (very much like in the case of the squared speed of sound, $V_s^2=dP/d\epsilon$). Below the critical temperature our results show a bump which is also observed in the usual \lsm, Polyakov \lsm and lattice studies (see Ref.\cite{china}). For high values of $T$ the quantity $w=P/\epsilon$ converges to approximately 1/3 as expected. 

\begin{figure}[tbh]
\vspace{0.5cm} 
\epsfig{figure=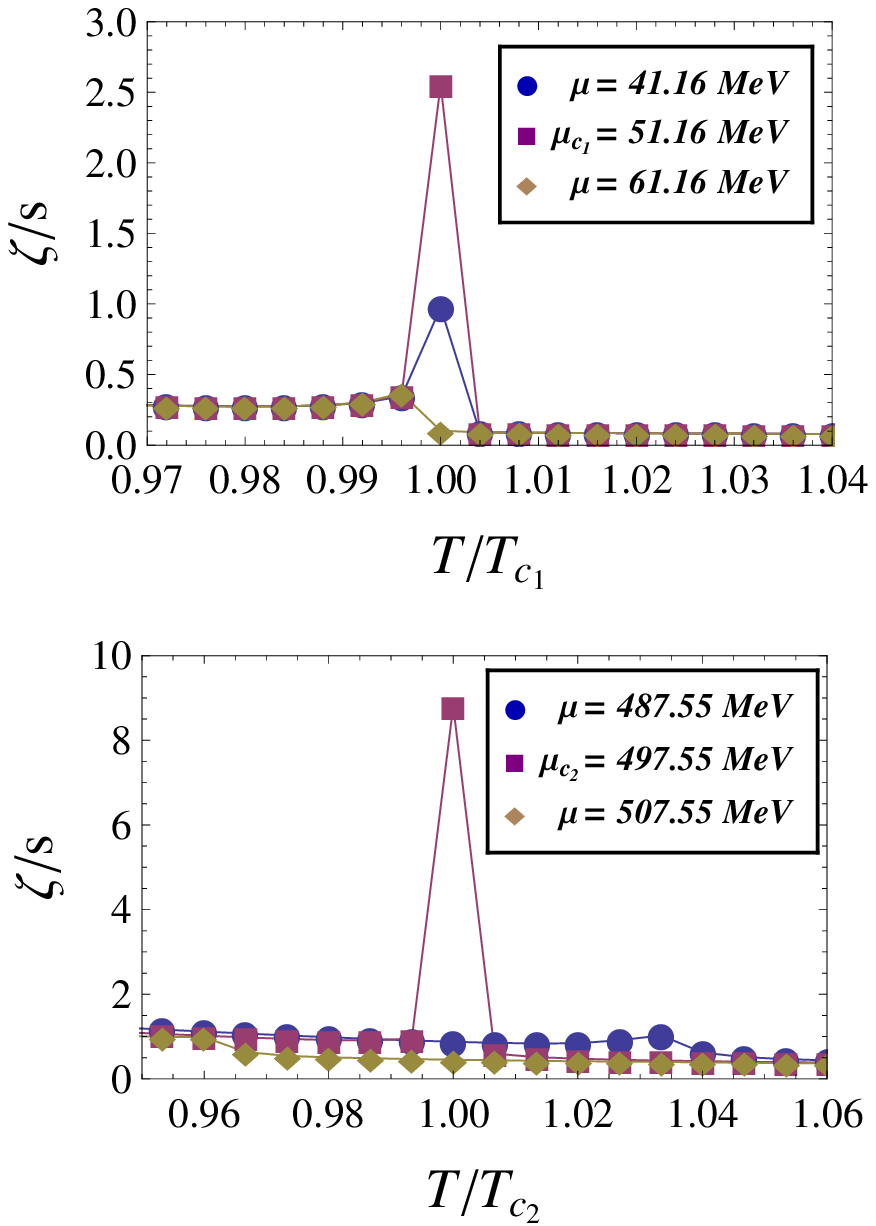,angle=0,width=7cm}
\caption{Bulk viscosity over entropy density, $\zeta/s$, as a function of the  temperature for the critical point $c_1$ (top
panel) and for $c_2$ (bottom panel). Chemical potential above and below the critical points values are also shown for reference. Here, $G \Lambda^2 =3.7$ and $m_c=0.1$.}
\label{zetas}
\end{figure}

The bulk viscosity, $\zeta$, is an intrinsic dynamical quantity which, however, can be
expressed in terms of the static thermodynamical quantities derived from the
free energy as~\cite{karsch}

\begin{equation}
\zeta = \frac{1}{9 \omega_0} \left[ T^5 \frac{\partial}{\partial T}
  \frac{(\epsilon-3 P)}{T^4} + 16 | \epsilon_0 | \right]\;,
\label{bulk}
\end{equation}
where $\omega_0$ is a scale which will be set to $\Lambda$, as in Ref.\cite {prc}, while $\epsilon_0$ represents the vacuum part of the energy density. Since $\zeta$ is proportional to the specific heat, $C_v$, the bulk viscosity over entropy density, $\zeta/s$, behaves as $1/V_s^2$ near $T_c$ in this approximation and peaks at the critical end points as indeed shown by our Fig. \ref {zetas} where the divergences at ($T_{c_1},\mu_{c_1}$) and at  ($T_{c_2},\mu_{c_2}$) indicate that the energy density has a sudden change at the critical points typical of a first order transition. For our purposes this is an interesting quantity since it has been pointed out that one can distinguish whether the system experiences a first order phase transition or a crossover from observables which are sensitive to the bulk viscosity in RHIC type of experiments. On expects that a sharp rise of bulk viscosity near a phase transition induces an instability in the hydrodynamic flow of the plasma, and this mode will  blow up tearing the system into droplets \cite{ china}.

Let us now examine numerically the behavior of the quark susceptibility, $\chi_{\q}$, as well as of the chiral susceptibility, $\chi_m$, near the two critical points. These quantities are respectively given by 

\begin{equation}
\chi_{\q} = \frac{\partial \rho_{\q}}{\partial \mu}\;\;,
\label{chiq_1}
\end{equation}
and
\begin{equation}
\chi_m = \frac{\partial \rho_s}{\partial m_c}\;\;.
\label{chiq_2}
\end{equation}

\begin{figure}[tbh]
\vspace{0.5cm} 
\epsfig{figure=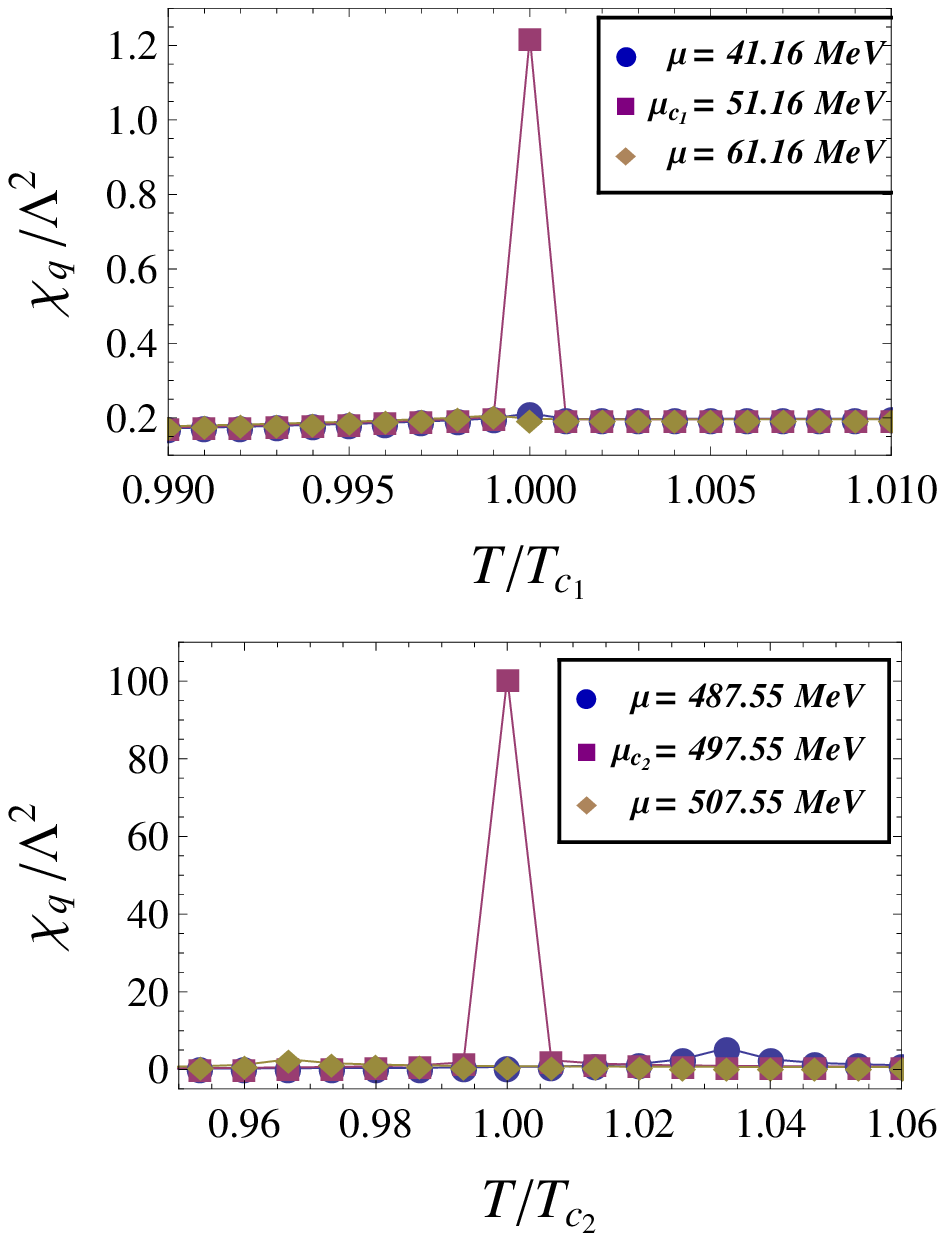,angle=0,width=7cm}
\caption{Normalized quark susceptibility, $\chi_{\q}/\Lambda^2$, for the critical point $c_1$ (top panel) and for $c_2$ (bottom panel). Chemical potentials above 
and below the critical points values are also shown for ref reference. Here, $G \Lambda^2 =3.7$ and $m_c=0.1$.}
\label{chiq}
\end{figure}

\begin{figure}[tbh]
\vspace{0.5cm} 
\epsfig{figure=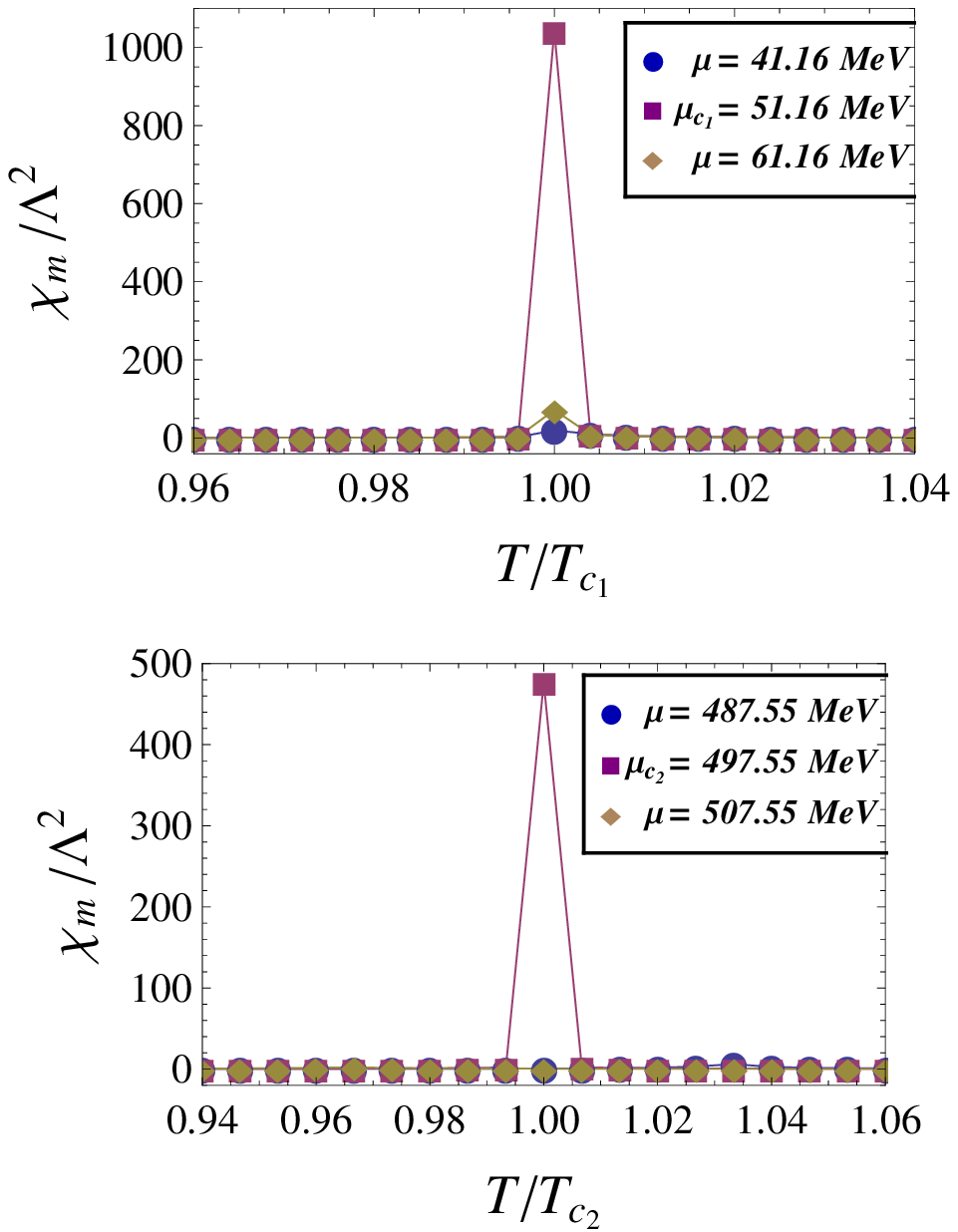,angle=0,width=7cm}
\caption{Normalized  chiral susceptibility, $\chi_{m}/\Lambda^2$, for the critical point $c_1$ (top
panel) and for $c_2$ (bottom panel). Chemical potentials above and below the 
critical points values are also shown for ref reference. Here, $G \Lambda^2 =3.7$ and $m_c=0.1$.}
\label{chim}
\end{figure}

Figures \ref {chiq} and  \ref {chim}  show $\chi_{\q}$ and $\chi_m$ respectively as functions of $T$ for relevant values of $\mu$. As expected, one observes that these two quantities peak for both $c_1$ and $c_2$. However, for $\chi_{\q}$, the magnitude of the peak associated with $c_2$ seems to be much larger while for $\chi_m$ the difference is not so dramatic. In principle these results could be interpreted as showing that the quark density, $\rho_{\q}$, plays a very minor role within $c_1$  while $\rho_s$ seems to dominate that critical point.  At the same time, the liquid-gas type of critical point, $c_2$, seems to receive contributions from both types of density in accordance with   our results for the \lsm.

\begin{figure}[tbh]
\vspace{0.5cm} 
\epsfig{figure=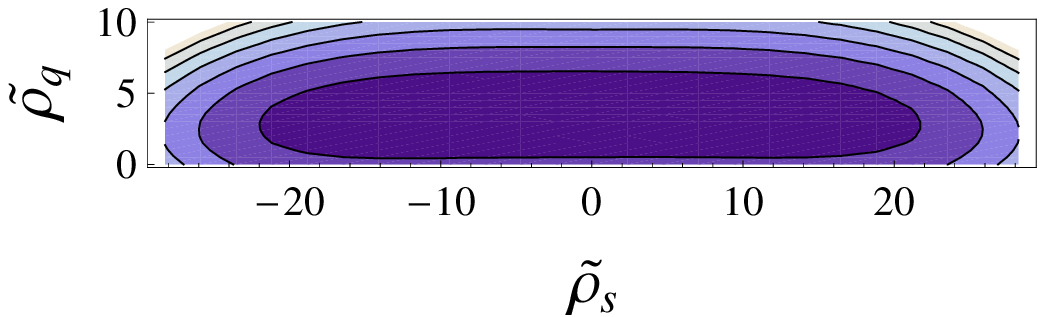,angle=0,width=6.7cm}

\epsfig{figure=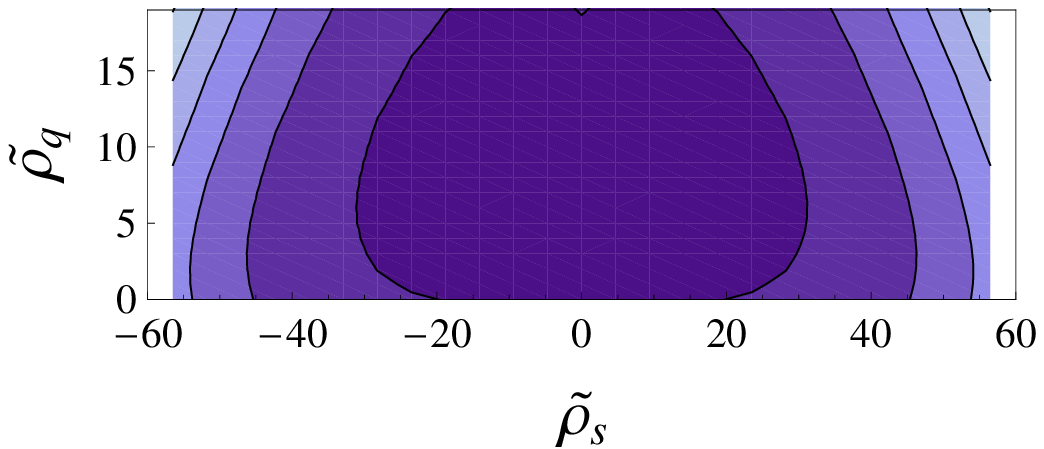,angle=0,width=6.7cm}

\epsfig{figure=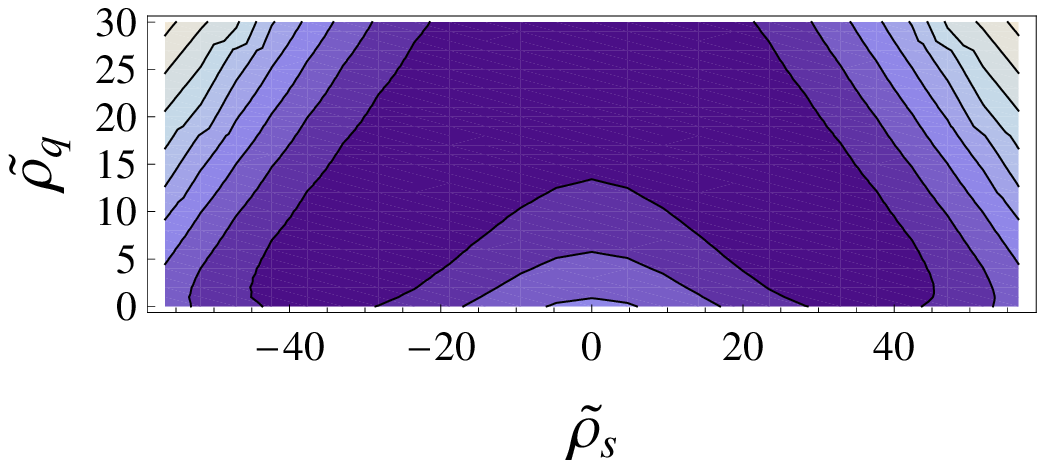,angle=0,width=6.7cm}
\caption{The Legendre transformed effective potential, $ { {{\cal V}_{eff}}}(T,\mu,m_c;\rho_s, \rho_{\q})$, projected in the dimensionless plane ${\tilde \rho}_s-{\tilde \rho}_q$ plane, where ${\tilde \rho} = \rho/(10^6 {\rm MeV}^3)$ in both cases. The top panel shows this quantity at $c_1$, the middle panel shows it at $T=240 \, {\rm MeV},\mu=150 \, {\rm MeV}$ where a cross over takes place. The bottom panel shows the contour plot at $c_2$ }
\label{contour}
\end{figure}

These findings can be further appreciated if  one investigates the free energy in terms of the two ordering densities, $\rho_s$ and $\rho_{\q}$, by using the techniques of 
Fujii and Ohtani~\cite {fujii} to Legendre transform ${U_{eff}}(T,\mu,m_c)$ to $ { {{\cal V }_{eff}}}(T,\mu,m_c;\rho_s, \rho_{\q})$. The result of this type of manipulation is shown in Fig.~\ref  {contour} for the two critical points, $c_1$ and $c_2$, as well as at an intermediate point ($T=240 \, {\rm MeV},\mu=150 \, {\rm MeV}$) where a cross over takes place. The contour plots displayed by  Fig. \ref {contour} indicate  that $c_1$ is indeed dominated by the scalar interaction, $\rho_s$, while the quark (vector) density also plays an important  role at $c_2$ in accordance with the covariance matrix results for the \lsm.

\begin{figure}[h]
\vspace{0.5cm} 
\epsfig{figure=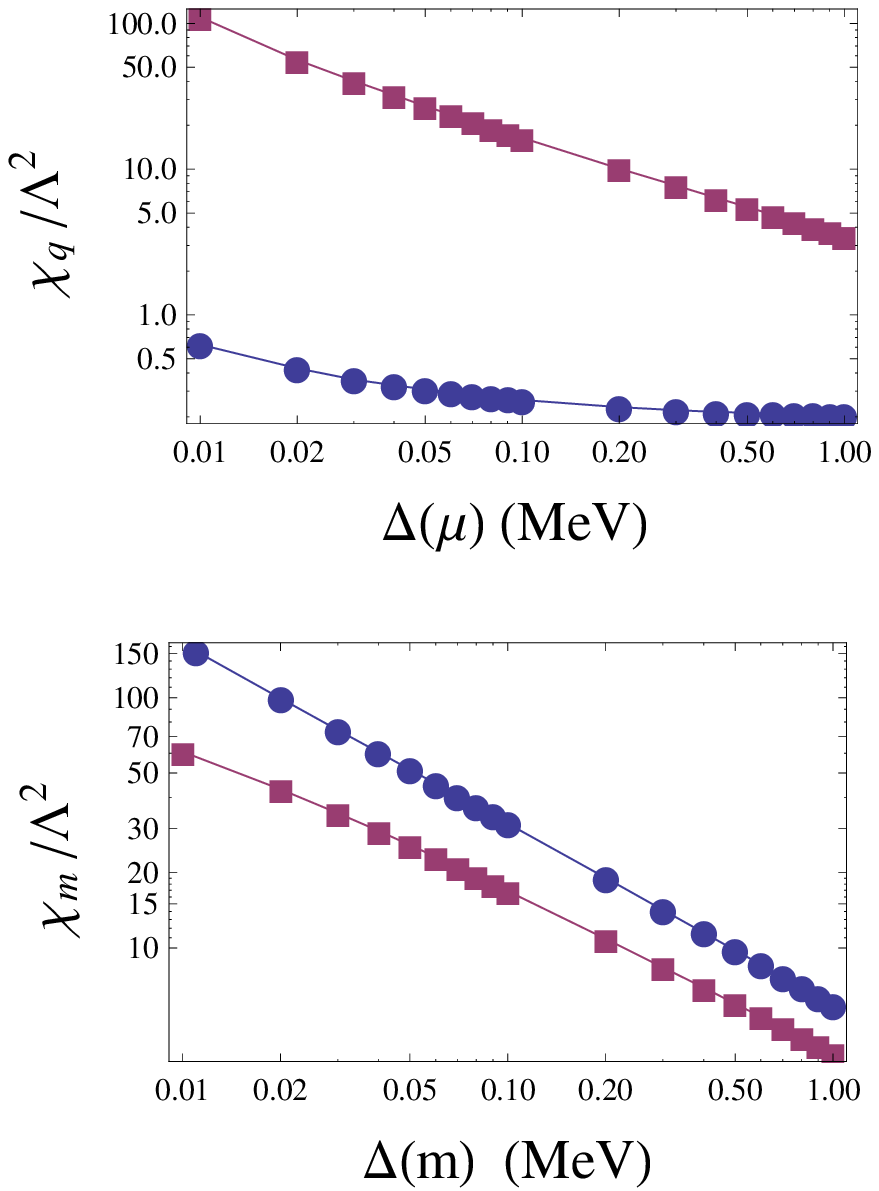,angle=0,width=6.7cm}
\caption{(color on line) Top panel: Logarithmic plot of the dimensionless $ \chi_{\q}/\Lambda^2$ as a function of $ \Delta(\mu)= |\mu - \mu_{c_2(c_1)}|$ approached from $\mu < \mu_{c_2(c_1)}$. The dots correspond to $c_1$ and the diamonds to $c_2$. Bottom panel: Same type of plot for the dimensionless $\chi_m/\Lambda^2$ as a function of $\Delta(m)= |m - m_c|$ approached from $m > m_c$.}
\label{ce}
\end{figure}

Finally, the data used in the coexistence phase diagram $T-\rho_B$  allow us to evaluate the critical exponent $\beta$ defined as
$  |\rho_{\q}^+ - \rho_{\q}^-| \propto   |T - T_E|^\beta$ where $E$ represents $c_1$ or $c_2$ while $\rho^+_{\q}$ and $\rho^-_{\q}$ represent the two corresponding densities for a given temperature. 
Having the quark number susceptibility allows for the numerical evaluation of the critical exponent $\epsilon$,  $ \chi_{\q} \propto  |\mu - \mu_E|^{- \epsilon}$, for which one may define a chiral counterpart, $\epsilon_m$, given by  $ \chi_{m} =  |m_c - m|^{- \epsilon_m}$. This procedure is 
illustrated by the top panel of Fig. \ref{ce} in which $\epsilon$ is obtained by approaching the critical point in a path parallel to the $\mu$ axis from $\mu < \mu_{c_1(c_2)}$. The values $\epsilon \simeq 0.64$ for $c_2$ is close to the one obtained in Ref.\cite {pedrosu2} while for $c_1$ the value is $\epsilon \simeq 0.50$. The bottom panel of Fig. \ref{ce} shows the same procedure for $\epsilon_m$ with the critical points being approached from the $m$ ``wing'' of the $T-\mu$ plane,  with $m>m_c$.
Interestingly enough the numerical values get approximately inverted, when compared to $\epsilon$, and one gets $\epsilon_m \simeq 0.49$ for $c_2$  and $\epsilon_m \simeq 0.63$ for $c_1$.
A similar type of procedure gives $\beta \simeq 0.48$ for $c_1$ and $\beta \simeq 0.35$ for $c_2$.  Having $\beta$ and $\epsilon$ allows us to determine $\delta$, $\gamma$ and $\alpha$ using  $\epsilon= 1- 1/\delta$, $\gamma=\beta(\delta-1)$ and $\alpha+2\beta+\gamma=2$. The values for the remaining exponents  have been obtained after approximating our numbers for $\beta$ and $\epsilon$ by the ratios  shown in Table I. Note that although our numerical estimates are very crude, since we do not cover many orders of magnitude nor try possible different paths leading to the critical points, they support our previous discussion regarding the nature of the two critical points found in this work. In particular, the values for $\beta$ listed in Table I  support the liquid-gas character of $c_2$ \cite {ma}. For this critical point also the exponent $\epsilon$, associated with $\chi_{\q}$, is greater than the one associated with the critical point $c_1$. Note also that the values of $\alpha$ obtained by using our approximate ratios for $\beta$ and $\epsilon$ in the scaling  relations show that $\alpha=\epsilon$ for both critical points which is consistent with the universal arguments presented in Refs. \cite{hatta02,universal} since it is expected that $\chi_{\q}$ and $C_v$ should be the same near the critical points.
A detailed and highly precise determination of the associated critical exponents is beyond the scope of the present application and the interested reader is referred to Refs. \cite {hatta02,pedrosu2,pedro}. 
\begin{table}[th]
\caption{Approximate ratios for the critical exponents associated with $c_1$ and $c_2$ in the NJL model. The values in between brackets are the results which have been numerically  obtained.}
\centering
\begin{tabular}{ccccccc}
\hline
 CEP & $\alpha$&$\beta$ &$\gamma$&$\delta$& $\epsilon$ &  $\epsilon_m$  \\ 
\hline  
\hspace{1.5cm} & \hspace{1.5cm} & \hspace{1.5cm} &  \hspace{1.5cm} & \hspace{1.5cm} & \hspace{1.5cm} &  \hspace{1.5cm} \\ 
$c_1$ &1/2&1/2 (0.48)&1/2&2&1/2 (0.50)&2/3 (0.63)\\ 
$c_2$ &2/3 & 1/3 (0.35)& 2/3&3&2/3 (0.64)&1/2 (0.49)\\ 
\hline
\end{tabular}
\label{table-results-p4}
\end{table}

\subsection{Back-bending in the $\mu-m_c$ plane}

\begin{figure}[tbh]
\vspace{0.5cm} 
\epsfig{figure=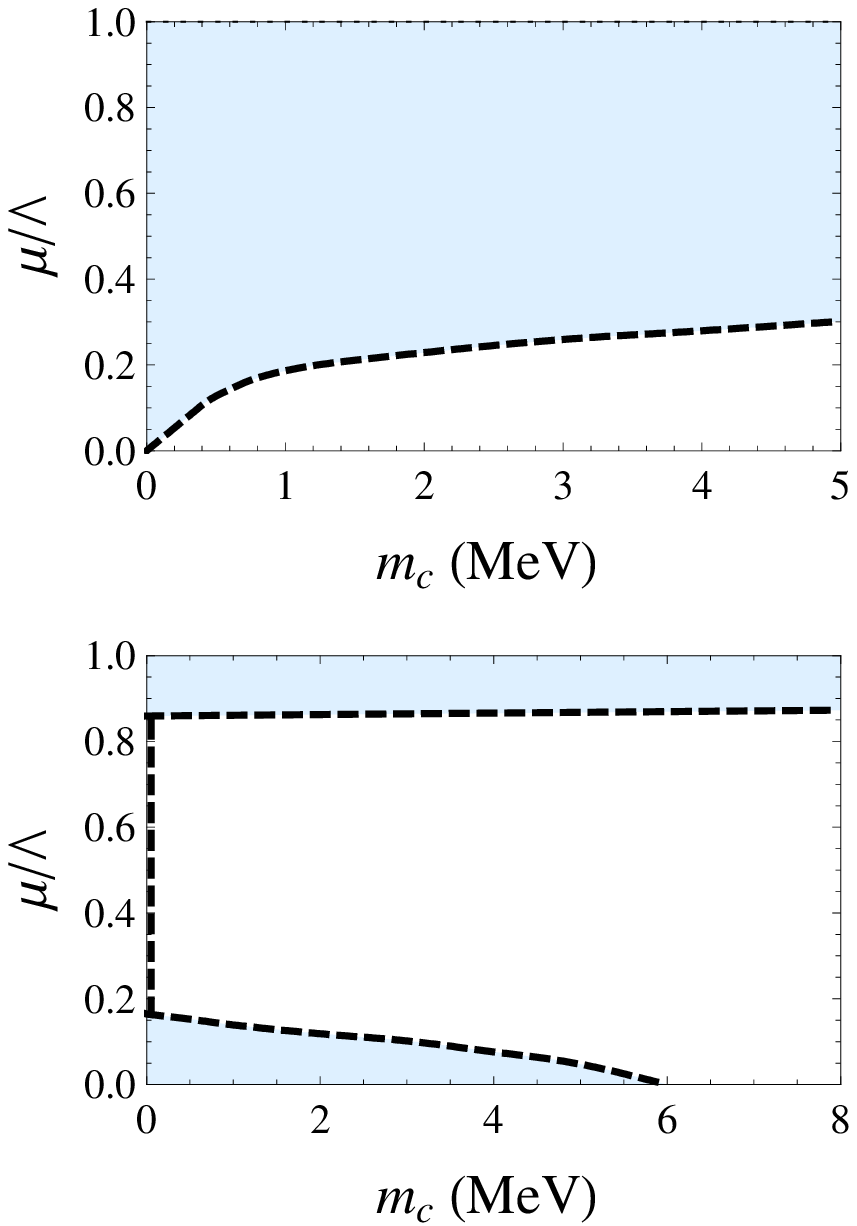,angle=0,width=7cm}
\caption{The $\mu$  versus $m_c$ plane showing the second order transition boundary (dashed lines) associated with the 
critical point(s) for the NJL. 
The shadowed regions correspond to first order transitions and the white region represents the cross over region. 
The top panel corresponds to MFT and shows only a single
first order branch with one critical point. The bottom panel corresponds to OPT and shows the possibility of two first order branches associated with two 
distinct critical points. Both results are for $G\Lambda^2=3.98$.
}
\label{mcmu}
\end{figure}
It will be interesting to explicitly show that, at least  from a qualitative perspective,  our model 
predictions can be relevant for the   lattice results obtained  by de Forcrand and Philipsen 
\cite{deForcrand:2006pv} who observed a shrinkage of the first order transition region when considering 
higher values of $\mu$.
For completeness let us also compare the OPT with the MFA results when the model parameters are tuned 
so that the latter approximation also generates a first order transition at $\mu=0$. 
Within the NJL model, the MFA can predict a first order transition at $\mu=0$ and 
for $\Lambda=590\, {\rm MeV}$ when  the higher coupling, $G \Lambda^2=3.98$, is used with $m_c=0.1 \, {\rm MeV}$. In this case, the MFA also 
predicts $f_\pi \simeq 93 \,{\rm MeV}$, $m_\pi \simeq 20 \,{\rm MeV}$, 
and $-\langle {\bar \psi \psi} \rangle^{1/3} \simeq 263.5 \, {\rm MeV}$, and $m_q^0 \simeq 814 \, {\rm MeV}$ while, 
for the same set of parameters, the OPT predicts 
$f_\pi \simeq 90.68 \,{\rm MeV}$, $m_\pi \simeq 20.73 \,{\rm MeV}$, and 
$-\langle {\bar \psi} \psi \rangle^{1/3} \simeq 265 \, {\rm MeV}$, and $m_q^0 \simeq 856 \, {\rm MeV}$. 

In the $\mu-m_c$ plane the NJL model then generates Fig.~\ref  {mcmu} in which the top panel, corresponding to the MFA, shows only a one branch first order line. 
When the OPT is applied to the NJL with the stronger $G \Lambda^2=3.98$ considered in this subsection   one observes a first 
order line for $m_c < 3.5 \, {\rm MeV}$ which roughly corresponds to $m_\pi = 122 \, {\rm MeV}$. 
Starting at the 
point $m_c=3.5 \, {\rm MeV}, \mu=0$  one may follow a second order transition line which 
goes left touching the $\mu$ axis at $\mu \simeq 70\, {\rm MeV}$ and then going up to $\mu \simeq 350\, {\rm MeV}$ where it bends back to the right hand side 
for finite values of $m_c$. The situation is illustrated by the bottom panel of figure \ref {mcmu}.

Therefore, when one goes beyond MFA, the NJL also predicts a bending back behavior of the critical line in a way which 
is consistent with the negative curvature of the lattice simulations of Refs.~\cite{deForcrand:2006pv}. This  result also supports Fukushima's 
suggestion~\cite {Fukushima:2008is} regarding the eventual back bending.   Although this author has considered the SU(3) 
version of the NJL (being able to reproduce a critical surface in the $m_{u,d}-m_s-\mu$ space)    
with standard parametrization it is interesting to note that the bending was obtained by adding a vector 
interaction which generated a $- G_V \rho_{\q}^2$ contribution to the pressure. By looking at our 
Eq. (\ref  {omegarhos}) one sees that the OPT also brings a term like $-G/(N_c N_f) \rho_{\q}^2$ to $P=-\Omega$. Finally, note that a back bending behavior is also implied by our \lsm results if one considers that $c_2$ and $c_2^\prime$ almost coincide.

\section{Conclusions}

In recent publications~\cite{2tri,BoKa} it was shown that for small values of the vacuum pion mass ($\lesssim 50$~MeV) the 
phase diagram of the L$\sigma$M has two distinct first order lines terminating in two critical points.
One, as predicted by most models, starts at finite $\mu$ and $T=0$, whereas the other is an unusual 
first order line which starts at high $T$ and $\mu=0$. In between the critical endpoints 
(whose exact location depends on the vacuum pion mass) of these two 
lines there is a crossover transition. At the origin of this behavior, that is not observed in MFA, are the 
thermal fluctuations of the mesonic fields, that have been taken into account adopting a self consistent 
method first proposed in~\cite{Mocsy:2004ab}. Inspired by this finding we have performed a careful analysis 
of the \lsm with the aim of understanding in deeper details the nature of the critical points. 
At the same time, we have considered another popular effective quark model, the NJL model, to investigate 
under which conditions a phase diagram with at least two distinct critical points could be reproduced.  

The analysis of the \lsm has been performed by using the same method of Ref.~\cite{2tri} and our results
agree with the ones obtained therein: the unusual 
first order line which starts at high $T$ and $\mu=0$ ending in the ``new'' critical point ($c_1$) was also found.
At the same time, a closer look in the vicinity of the ``usual'' critical endpoint revealed that, 
right before to end, the first order line that starts at finite 
$\mu$ and $T=0$ bifurcates in {\em two} critical endpoints ($c_2$ and $c_2'$) separated by few 
MeV in $T$ and $\mu$.

In addition to the phase diagram we have studied susceptibilities and correlations of the net-quark number density, 
the entropy density, and the scalar density. 
Our results for the covariant matrix show that for $c_1$ the dominant fluctuations 
are given by the scalar density and the entropy density, whereas the quark number density plays only a minor role.
For the two critical points $c_2$ and $c_2'$, instead, also the fluctuations of the quark number density 
are important. Their contribution becomes larger and larger as the vacuum pion mass is increased,  
probably also due to the fact that $c_2$ and $c_2'$ move towards higher values of $\mu$. 
Despite their vicinity in the phase diagram, $c_2$ and $c_2'$ seem to exhibit distinct features.
The main difference being the fluctuations of the scalar density, that are more important 
in $c_2'$ than in $c_2$.

We have then considered the NJL thermodynamical potential, recently 
evaluated with the OPT~\cite {prc}, with parameter values which simulate small pion masses in order to generate a $T-\mu$ phase diagram 
where the existence of  two critical points, $c_1$ and $c_2$, separated by a cross over region 
has  been observed. 
This type of phase diagram is similar to the one found, in the temperature-magnetic field plane, for the compressible metamagnetic Ising model \cite {wagner}

We have performed an extensive thermodynamical analysis in order to find the essential physical features that distinguish both critical points concluding, in agreement with the \lsm case, that the usual one ($c_2$), located at intermediate values of $T$ and $\mu$ has a more hydrodynamical character with the quark number density playing an essential role. This density   has little influence at lower values of $\mu$ where the new, unusual, critical point $c_1$ region is dominated by the scalar density.
 Although we have not 
considered a vector term in the version of the NJL model considered here it is rather interesting to note 
that the OPT pressure has the term  $-G/(2N_f N_c)(2\rho_{\q}^2-\rho_s^2)$ which by being $1/N_c$ suppressed 
does not contribute to the MFA which completely misses out the possible existence of $c_1$ and the 
associated first order line.

From these results it is clear that the inclusion of contributions beyond MFA can have dramatic 
consequences on the phase diagram of quark models and cannot be neglected.
With the combined analysis of the \lsm and the NJL model, we have shown that the appearance 
of multiple critical points is not an uncommon feature of effective quark models, even without 
explicitly introducing a vector interaction as was done in~\cite{Fukushima:2008is}. 
Before exporting these notions to QCD, however, some caution is advisable.
As has been very recently shown in Ref.~\cite{Skokov:2010sf}, based on the analysis of the \lsm in MFA,
the inclusion of vacuum fluctuations can change the transition in the chiral limit at $\mu=0$ from 
first to second order so that this model would behave as the NJL model in the same regime. However, our results for the latter suggest that  the consideration of vacuum contributions  should not
 influence the appearance of the critical point $c_1$, at least qualitatively, when  an appropriate tuning of model parameters is carried out beyond the MFA. 
In the minimal version of the NJL model considered here $c_1$ appears only if the coupling is 
larger than usual (see discussion in sec.~\ref{nonstandard})  while a sharp cut off is introduced only in the divergent integrals. However, having a large parameter space, more sophisticated versions of the model could also allow for the appearance of a critical point point like $c_1$ generating, at the same time, more plausible values for the vacuum effective quark mass.
Concerning the other two critical points found in the \lsm ($c_2$ and $c_2'$), instead, 
vacuum fermion loops are not expected to modify dramatically the qualitative behavior of the 
model, especially for values of the vacuum pion mass close to the physical one~\cite{Mocsy:2004ab}. 
The presence of $c_2'$ in the \lsm with thermal fluctuations, however, has not been confirmed  by the OPT 
analysis of the NJL model. This might be due to various reasons including the fact that the two models have 
been evaluated under different approximations, the fact that the NJL does not have true mesonic degrees of 
freedom, etc. The clarification of these points calls  for  further investigations.  
Nevertheless, our results allow us to conclude that it is possible to use these effective models to generate 
metamagnetic like phase diagrams at the expense of using non standard parameters for $m_c$ and $m_\pi$ which 
weakly break chiral symmetry in approximations which go beyond the MFA. Having  two critical points, it then 
becomes possible to observe the shrinkage of first order region, in the $\mu-m_c(m_\pi)$ plane,  
as one considers higher values of $\mu$ so that  model predictions could be conciliated with  the lattice 
results by de Forcrand and Philipsen \cite{deForcrand:2006pv}. Then, as we have shown, the first order region 
should increase again eventually intercepting the physical mass point as argued in Ref.\cite {BoKa}. 
The ``chiral'' like first order transition observed in the line associated with $c_1$ seems to be much 
softer and the coexistence regions much smaller  than in the traditional ``liquid-gas'' type of line 
associated with the $c_2$ which perhaps would make it harder to be detected in  lattice QCD calculations
(see for example ref.~\cite{Allton:2005gk}). Although quantities such as the trace anomaly, EoS 
parameter, bulk viscosity and susceptibilities 
display the expected  behavior associated with first order phase transitions it seems that the 
different critical points belong to distinct universality classes. 

\section*{Acknowledgments}

M.~B.~P.  thanks the Nuclear Theory Group at LBNL for the hospitality during the sabbatical year.  We thank J.-L. Kneur, R. Ramos, I. N. Mishustin,  W. Figueiredo,  P. Costa, Y. Hatta, H. Hansen, and A. Delfino for discussions.
This work was supported  by the Director, Office of Energy
Research, Office of High Energy and Nuclear Physics, Divisions of
Nuclear Physics, of the U.S. Department of Energy under Contract No.
DE-AC02-05CH11231, by the Helmholtz International Center for FAIR within the
framework of the LOEWE program (Landesoffensive zur Entwicklung Wissenschaftlich-\"Okonomischer Exzellenz) launched by the
State of Hesse, and by Coordena\c c\~{a}o de Aperfei\c{c}oamento de Pessoal de N\'{\i}vel  Superior
(CAPES, Brazil).

\end{document}